\pdfoutput=1 

\documentclass[onecolumn,notitlepage,showpacs,preprintnumbers]{revtex4-1}

\usepackage{verbatim} 
\usepackage{epsfig}
\usepackage{amssymb} 
\usepackage{amsmath}

\usepackage{mathtools}

\usepackage{lineno}

\usepackage{tikz}	
\usepackage{etoolbox}

\usepackage[caption=false]{subfig} 
\usepackage{dcolumn}
\usepackage{bm}

\begin{document}

\newcommand*{\hwplotB}{\raisebox{3pt}{\tikz{\draw[red,dashed,line width=3.2pt](0,0) -- 
(5mm,0);}}}

\newrobustcmd*{\mydiamond}[1]{\tikz{\filldraw[black,fill=#1] (0,0) -- (0.2cm,0.2cm) --
(0.4cm,0) -- (0.2cm,-0.2cm);}}

\newrobustcmd*{\mytriangleright}[1]{\tikz{\filldraw[black,fill=#1] (0,0.2cm) -- 
(0.3cm,0) -- (0,-0.2cm);}}

\newrobustcmd*{\mytriangleleft}[1]{\tikz{\filldraw[black,fill=#1] (0,0.2cm) -- 
(-0.3cm,0) -- (0,-0.2cm);}}
\definecolor{Blue}{cmyk}{1.,1.,0,0}




\title{High pressures in room evacuation processes and a first 
approach to the dynamics around unconscious pedestrians}

\author{F.E.~Cornes}
 \affiliation{Departamento de F\'\i sica, Facultad de Ciencias 
Exactas y Naturales, Universidad de Buenos Aires,\\
 Pabell\'on I, Ciudad Universitaria, 1428 Buenos Aires, Argentina.}
\author{G.A.~Frank}
 \affiliation{Unidad de Investigaci\'on y Desarrollo de las 
Ingenier\'\i as, Universidad Tecnol\'ogica Nacional, Facultad Regional Buenos 
Aires, Av. Medrano 951, 
 1179 Buenos Aires, Argentina.}
\author{C.O.~Dorso}%
 \email{codorso@df.uba.ar}
\affiliation{Departamento de F\'\i sica, Facultad de Ciencias 
Exactas y Naturales, Universidad de Buenos Aires,}%
 \affiliation{Instituto de F\'\i sica de Buenos Aires,
 Pabell\'on I, Ciudad Universitaria, 1428 Buenos Aires, Argentina.}%
\date{\today}




\begin{abstract}
Clogging raises as the principal phenomenon during many evacuation processes 
of pedestrians in a panic situation. As people push to escape from danger, 
compression forces may increase to harming levels. Many individuals might fall 
down, while others will try to dodge the fallen people, or, simply pass through 
them. We studied the dynamics of the crowd for these situations, in the context 
of the ``social force model''. We modeled the unconscious (fallen) pedestrians 
as inanimate bodies that can be dodged (or not) by the surrounding individuals. 
We found that new morphological structures appear along the evacuating crowd. 
Under specific conditions, these structures may enhance the evacuation 
performance. The pedestrian's willings for either dodging or passing through 
the unconscious individuals play a relevant role in the overall evacuation 
performance.  
\end{abstract}









\maketitle


\section{\label{introduction}Introduction}

Recent investigation on panic evacuation has achieved a wide variety of 
scenarios and behavioural patterns \cite{frank2015,frank2016}. But, to our 
knowledge, the research has not dug deeply enough into the consequences of 
injuries and  unconsciousness. \\

History acknowledges many fatalities during stampedes. Unfortunately, 
such kind of disasters have increased in frequency because the 
number and size of massive events (music festivals, sports events, 
etc.) has become larger \cite{fruin}.  An inspection of the Crowdsafe 
Database$^{TM}$ through 1992 to 2002 shows a correlation between the number of 
concerts and festival events, and the number of injuries 
\cite{crowdsafe}. Specially sorrowful are the incidents occurred in 
the nightclubs  \textit{The Station} (Rhode Island, 2003) and 
\textit{Croma\~n\'on} (Buenos Aires, 2004) where 100 and 194 people lost their 
lives, 
respectively.  \\

The overcrowding is one of the principal causes of injury or death while people 
try to escape under panic. Deaths may happen because of \textit{trampling} or 
\textit{compression due to crush}. The former occurs when someone falls in a 
high dense crowd, not being able to stand again due to the movement of the 
others, unaware of the fallen pedestrian. This produces a continue trampling 
that finally kills the individual \cite{lee}. \\

Compression due to crush is the other cause of death. This effect appears in 
high dense crowds, preventing the free movement of the pedestrians. If the 
pressures in the crowd become extremely high, each time an individual breaths 
out, the pressure restricts the inhalation of the next breath. Thus, 
compression due to crush causes asphyxia on the individual, evolving to 
unconsciousness or death after some time \cite{gill}. Further information on 
fatal consequences by asphyxia can be found in Ref.~\cite{lee}. \\

A brief review of the basic ``social force model'' can be found in 
Section~\ref{sec:The SocialForceModel}. We include in 
Section~\ref{Adjustmentstothemodel} an upgrade of the basic model that makes 
possible to achieve compressional injuries.  \\

In Section~\ref{experimentaldata} we will present experimental data on the 
injury threshold due to compression. A simple model on the human torso will be 
examined for further simulations (see Section~\ref{simulations}). \\  

All the results of our investigations are presented in Section~\ref{results}. 
The corresponding conclusions are summarized in Section~\ref{conclusions}. \\

\section{\label{background}Background}

\subsection{The Social Force Model}
\label{sec:The SocialForceModel}

The ``social force model'' (SFM) proposed by Helbing and co-workers 
\cite{helbing} is a generalized force model, including socio-psychological 
forces, as well as ``physical'' forces like friction. These forces enter the 
Newton equation as follows.   

\begin{equation}
m_i\,\displaystyle\frac{d\mathbf{v}^{(i)}}{dt}=\mathbf{f}_d^{(i)}
+\displaystyle\sum_{j=1}^{N}\displaystyle\mathbf{f}_s^{(ij)}
+\displaystyle\sum_ {
j=1}^{N}\mathbf{f}_g^{(ij)}\label{eq_mov}
\end{equation}

\noindent where the $i,j$ subscripts correspond to any pedestrian in the 
crowd. $\mathbf{v}^{(i)}(t)$ means the current velocity of the pedestrian  
$(i)$, while $\mathbf{f}_d$ and $\mathbf{f}_s$ are the socio-psychological 
forces acting on him (her). $\mathbf{f}_g$ is the friction or granular force. \\

$\mathbf{f}_d(t)$ and $\mathbf{f}_s(t)$ are essentially different. The former 
corresponds to the ``desire force'', that is, the pedestrians own willings to 
move towards a desired position. The latter corresponds to the ``social 
force'', meaning the tendency of the pedestrians to preserve their 
\emph{private sphere}. The ``social force'' prevents the pedestrians from 
getting too close to each other. \\

According to the anxiety state of the pedestrian, he (she) will accelerate (or 
decelerate) to reach any desired velocity $v_d$ that will make him (her) feel 
more comfortable. Thus, in the social force model, the desired force reads 
\cite{helbing}

\begin{equation}
        \mathbf{f}_d^ {(i)}(t) =  
m_i\,\displaystyle\frac{v_d^{(i)}\,\mathbf{e}_d^
{(i)}(t)-\mathbf{v}^{(i)}(t)}{\tau} \label{desired}
\end{equation}

\noindent where $m_i$ is the mass of the pedestrian $i$ and $\tau$ represents 
the relaxation time needed to reach his (her) desired velocity. 
$\mathbf{e}_d$ is the unit vector pointing to the target position. For 
simplicity, we assume that $v_d$ remains constant during an evacuation process, 
but $\mathbf{e}_d$ changes according to the current position of the 
pedestrian. Detailed values for $m_i$ and $\tau$ can be found in 
Refs.~\cite{helbing,frank2011}. \\

The \textit{private sphere} preservation corresponds to a repulsive feeling 
between the pedestrians, or, between pedestrians and the walls 
\cite{helbing,helbing1995}. These repulsive feelings become stronger as 
people get closer to each other (or to the walls). Thus, in the context of the 
social force model, this tendency is expressed  as \\

\begin{equation}
        \mathbf{f}_s^{(ij)} = A_i\,e^{(r_{ij}-d_{ij})/B_i}\mathbf{n}_{ij} 
        \label{social}
\end{equation}

\noindent where $(ij)$ represents any pedestrian-pedestrian pair, or 
pedestrian-wall pair. $A_i$ and $B_i$ are two fixed parameters (see 
Ref.~\cite{dorso2005}). The distance $r_{ij}=r_i+r_j$ is the sum of the 
pedestrians radius, while $d_{ij}$ is the distance between the center of mass 
of the pedestrians $i$ and $j$. $\mathbf{n}_{ij}$ means the unit vector in the 
$\vec{ji}$ direction. For the case of repulsive feelings with the walls, 
$d_{ij}$ corresponds to the shortest distance between the pedestrian and the 
wall, while $r_{ij}=r_i$ \cite{helbing,helbing1995}. \\


The granular force $\mathbf{f}_g$ included in Eq.~(\ref{eq_mov}) corresponds 
to the sliding friction between pedestrians in contact, or, between pedestrians 
in contact with the walls. The expression for this force is 

\begin{equation}
        \mathbf{f}_g^{(ij)} = 
\kappa\,(r_{ij}-d_{ij})\,\Theta(r_{ij}-d_{ij})\,\Delta
\mathbf{v}^{(ij)}\cdot\mathbf{t}_{ij} 
        \label{granular}
\end{equation}

\noindent where $\kappa$ is a fixed parameter. The function 
$\Theta(r_{ij}-d_{ij})$ is zero when its argument is negative (that is, 
$r_{ij}<d_{ij}$) and equals unity for any other case (Heaviside function). 
$\Delta\mathbf{v}^{(ij)}\cdot\mathbf{t}_{ij}$ represents the difference between 
the tangential velocities of the sliding bodies (or between the individual and 
the walls).   \\

\subsection{\label{Adjustmentstothemodel}Body compression in the Social Force 
Model}

\begin{figure}
\centering
\includegraphics[scale=0.45]{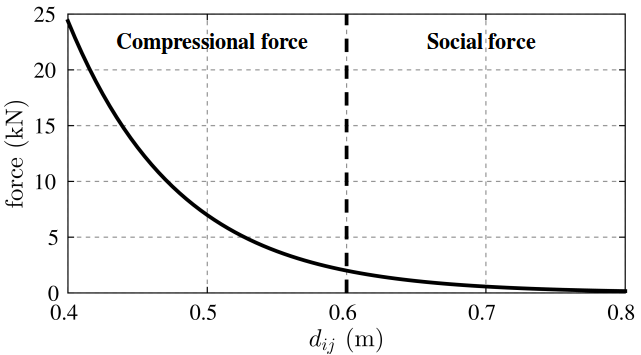}
\caption{\label{social_force_graph} Compressional force ($\mathbf{f}_c$) 
and social force ($\mathbf{f}_s$) as a function of the inter-pedestrian 
distance $d_{ij}$. The dashed line at 0.6~m corresponds to the meeting distance 
between $\mathbf{f}_c$ and $\mathbf{f}_s$.}
\end{figure}

Pedestrians that get in contact (or pedestrians that contact the walls) may 
experience some kind of body compression. The contacting force within the 
(basic) social force model is similar to Eq.~(\ref{social}), but for 
$d_{ij}<r_{ij}$. While the \textit{private sphere} preservation applies to 
repulsive feelings for $d_{ij}>r_{ij}$, the compressional force corresponds to 
a contact phenomenon. That is, the social and compressional forces may be 
represented as two piecewise functions \\

\begin{equation}
\left\{\begin{matrix}
\mathbf{f}^{(ij)}_c=A_{i}e^{(r_{ij}-d_{ij})/B_{i}}\mathbf{n}_{ij}
\,\Theta[r_{ij}-d_{ij}]
\\ 
\\ 
\mathbf{f}^{(ij)}_s=A_{i}e^{(r_{ij}-d_{ij})/B_{i}}\mathbf{n}_{ij}
\,\Theta[d_{ij}-r_{ij}]
\end{matrix}\right.
\label{eq_social_force_graph}
\end{equation} \\

\noindent where $\Theta$ is the Heaviside function. Thus, the social force 
$\mathbf{f}_s$ and the compressional force $\mathbf{f}_c$ share the same 
mathematical expression, despite that they have different meanings and apply to 
non-overlapping domains (see~Fig.~\ref{social_force_graph}). Notice that the 
movement equation (\ref{eq_mov}) needs no further modification because of this. 
\\

In order to capture the physical meaning of the compressional force 
$\mathbf{f}^{(ij)}_c$, we expand the expression in 
(\ref{eq_social_force_graph}) into powers of $r_{ij}-d_{ij}$ (within the 
domain $r_{ij}>d_{ij}$). The first order terms are as follows

\begin{equation}
\mathbf{f}^{(ij)}_c=A_i\,\mathbf{n}_{ij} +
\displaystyle\frac{A_i}{B_i}\,(r_{ ij }
-d_ { ij } )\,\mathbf{n}_{ij}+
\mathcal{O}[(r_{ij}-d_{ij})^2]\,\mathbf{n}_{ij}
\label{taylor}
\end{equation}

The first term on the right $A_i\,\mathbf{n}_{ij}$ corresponds to the repulsive 
feelings at $d_{ij}=r_{ij}$. The second term on the right resembles the Hooke 
law with elastic coefficient $A_i/B_i$. This law applies to small body 
compressions. The third term on the right, however, resembles the 
stiffness for large compressions. According to literature values, the elastic 
coefficient $A_i/B_i$ is $25000\,$N/m (see Refs.~\cite{helbing,frank2011}). In 
Section \ref{experimentaldata} we will show that this value is in agreement 
with experimental data for small body compressions. \\

\subsection{\label{Totaleffectiveforce}The effective compressional force}

We are interested in the forces causing body deformation in the front-back 
direction. This are actually the forces that may cause injury to the 
pedestrians. Therefore, we define the following ``effective'' compressional 
force\\

\begin{equation}
f^{(i)}_{e} =\sum_{j\neq i}^{N}\left | (\mathbf{f}_{c}^{(ij)}- 
A_{i}\mathbf{n}_{ij}) \cdot\mathbf{e}_{d}^{(i)} \right 
|\Theta[r_{ij}-d_{ij}]
\label{effective_force}
\end{equation} 

\noindent where $A_i\,\mathbf{n}_{ij}$ is the repulsive feeling at the 
contacting distance $d_{ij}=r_{ij}$. The inner product produces the projection 
of the compressional force (excluding the repulsive feelings at 
$d_{ij}=r_{ij}$) in the front-back direction. The bars means the modulus of the 
magnitude. The sum does not include the walls, since we assume that the 
desired direction $\mathbf{e}_{d}^{(i)}$ is tangential to the walls surface.  \\

\subsection{\label{pressure}The local pressure on the pedestrians}

The ``social pressure'' on a single pedestrian (say, $i$) is 
\cite{helbing}

\begin{equation}
p^{(i)} =\frac{1}{2\pi r_{i}} \sum_{j\neq 
i}^{N}[\mathbf{f}_{s}^{(ij)}+\mathbf{f}_{c}^{(ij)}]\cdot\mathbf{n}_{ij}
\label{phelbing}
\end{equation} 

The sum $\mathbf{f}_{s}^{(ij)}+\mathbf{f}_{c}^{(ij)}$ does not 
overlap, while it ensures continuity at $d_{ij}=r_{ij}$. Recall that both 
forces point from any individual $j$ to the individual $i$, and thus, the inner 
product is always positive. \\ 

Notice that Eq.~(\ref{phelbing}) holds either if the pedestrians are in 
contact or not. The feelings for preserving the \textit{private sphere} 
actuate as a ``social pressure'' that makes possible for the individuals to 
change their behavioural pattern when they come too close to each other or to 
the walls. The compressional force also changes the moving pattern of 
the pedestrian, if two pedestrians get in contact.  \\

\subsection{\label{passoverforce}The pass through force}

Pedestrians are capable of passing through other fallen individuals. This 
situation, however, can not be achieved by the basic ``social force model'' 
(SFM). According to Section~\ref{sec:The SocialForceModel}, repulsive feelings 
(\textit{i.e.} the social force) actuate on neighboring pedestrians. Only the 
individual's own desire (\textit{i.e.} desire force) can balance these 
feelings because the granular force is actually orthogonal to the 
inter-pedestrian direction. Therefore, it might happen that, although the 
pedestrian wants to accelerate towards the target position, he (she) will get 
stuck because of repulsion. \\

The dynamics for passing through fallen individuals require further extensions 
of the basic SFM. The pedestrians who decide to pass through fallen people 
experience some kind of desire, regardless of their own \textit{private sphere} 
preservation. Thus, the ``passing through'' process implies that the repulsive 
feelings (\textit{i.e.} the social force) between the moving pedestrian and the 
fallen one do not play a role. The relevant force acting on the moving 
pedestrian during this process seems to be his (her) desire to ``pass through'' 
the fallen individual. \\

Recall that the desire force $\mathbf{f}_d^{(i)}$ is related to the 
acceleration (deceleration) needed to reach any desired velocity $v_d$. However, 
the pedestrians who pass through fallen individuals may experience 
additional moving difficulties that might change their desired velocity with 
respect to the free moving desired one. The ``passing through'' 
desired velocity is therefore not expected to be the same as $v_d$. Actually, 
the acceleration (deceleration) time for the ``passing through'' process may 
be also different from the current relaxation time $\tau$.\\

We postulate as a \textit{working hypothesis} that a force actuates during the 
``passing through'' process that is similar to the desire force 
$\mathbf{f}_d^{(i)}$. But, since this force corresponds to the willings of 
``passing through'' a fallen pedestrian, instead of reaching the target 
position, its parameters should be different from those in 
Eq.~(\ref{desired}). 
The ``passing through'' force is defined as follows   

\begin{equation}
\mathbf{f}^{(i)}_{p}(t)=\textit{m}_i\frac{\textit{v}^{(i)}_p\mathbf{e
}^{(i)}_p(t)-\mathbf{v}_i(t)}{{\tau}'}
\label{fuerza_de_salto}
\end{equation}

Notice that $\mathbf{f}_p^{(i)}$ shares the same mathematical expression as the 
desired force (Eq.~\ref{desired}). But, now, the desired velocity has been  
replaced by a ``desired passing through'' velocity $v_p$, representing the 
slowing down with respect to $v_d$ due to the additional difficulties of the 
``passing through'' context. Besides, the relaxation time $\tau$ has also been 
replaced by the relaxation time ${\tau}'$ of the moving individual during the 
``passing through'' process. The passing through direction $\mathbf{e}_p$, 
however, is the same as the desired velocity 
($\mathbf{e}_p^{(i)}=\mathbf{e}_d^{(i)}$) 
since passing through fallen individuals can only take place if the latter is in 
the same path as the former. \\

Notice that $\mathbf{f}_d^{(i)}$ and $\mathbf{f}_p^{(i)}$ do not overlap in 
time. That is, the ``passing through'' force replaces the usual desired force 
whenever the moving pedestrian is in contact with the fallen pedestrian. 
This means that Eq.~(\ref{eq_mov}) needs no further modification. \\

\subsection{\label{human}Clustering structures}

Clusterization is responsible for the time delays during an evacuation process, 
as explained in Refs.~\cite{dorso2005,dorso2007}. Thus, a definition for
the clustering structures appearing during an evacuation process is required. 
We define a \textit{human cluster} as the group of pedestrians that for any 
member of the group (say, $i$) there exists at least another member belonging to 
the same group ($j$) in contact with the former. That is, 

\begin{equation}
 i\in\mathcal{G} \Leftrightarrow \exists j\in\mathcal{G}/d_{ij}<r_i+r_j
\end{equation}

\noindent where $\mathcal{G}$ corresponds to any set of individuals. \\

One or more human clusters may be responsible for blocking the way out of the 
room. The minimum set of human clusters that are able to block the way out of 
the room will be called \textit{blocking clusters}. If only one 
human cluster exists, we will call this blocking situation as a \textit{total 
blocking}. If more than one human cluster exists simultaneously, we will call
this situation a \textit{partial blocking}. \\

\section{\label{experimentaldata}Experimental data}

It was mentioned in Section~\ref{introduction} that compression due to crush 
may cause unconsciousness or death in an overcrowded environment. In a panic 
situation, where people push hard to get out, compression due to neighboring 
pedestrians can raise until certain injury limit. Although it is not 
possible to determine empirically this limit, a lower bound for the true injury 
level can still be established from the corresponding pain threshold. 
Additional data on compression from human cadavers or anesthetized animals is 
also available. \\

Table~\ref{parametros_modelo_social} resumes typical parameter values for the 
human body. Force thresholds were measured for quasi-static situations 
(that is, impact velocities less than 1~m/s). The elastic coefficients 
result from data fitting procedures into the (linear) Hooke's law. For further 
details see Ref.~\cite{gordon}. \\

\begin{table}
\caption{Experimental data for human body compression. Surface figures 
correspond to mean human values. The tolerance threshold was measured on the 
abdomen or sternum of the individuals, and lasted 1 second, according to 
Ref.~\cite{evans}. The death threshold corresponds to forces applied on the 
chest during 15 seconds. The elastic coefficients on the sternum correspond to 
deflections smaller than 25 and 38~mm, respectively. The elastic coefficient on 
the thorax corresponds to deflections smaller than 41~mm.}

\centering 
\begin{center}
\begin{tabular}{l@{\hspace{20mm}}c@{\hspace{20mm}}c@{\hspace{15mm}}c@{\hspace{
15mm } } l}
 \hline
 Magnitude & Symbol & Value & Units & Refs. \\
 \hline  
Mean body surface  & $S_B$ & 1.750 & m$^2$ & \cite{meunier}  \\
Mean torso surface & $S_T$ & 0.068 & m$^2$ & \cite{gordon,Huston} \\
Force tolerance threshold & $F_T$ & 276-356 & N & \cite{evans} \\ 
Force death threshold & $F_D$ & 6227 & N & \cite{hopkins} \\ 
Elastic coefficient on sternum & $k_s$ & 13.1-21.9 & kN/m & \cite{gordon} \\
Elastic coefficient on thorax & $k_t$  & 26.2 & kN/m & \cite{gordon} \\
\hline
\end{tabular}\label{parametros_modelo_social}
\end{center}

\end{table}

Data shown in Table~\ref{parametros_modelo_social} focuses on the front-back 
direction. Body compression on the chest is, indeed, the relevant one since it 
restricts inhalation on each breathing cycle. Forces applied on the left-right 
sides of the body do not play an important role for human survivability (see 
Ref.~\cite{gill}). Notice that the ``effective'' compressional force defined in 
Section~\ref{Totaleffectiveforce} resembles the chest compression only. \\  

According to Table~\ref{parametros_modelo_social}, the thorax seems to be 
somehow stiffer than the sternum. But, the upper bound for the sternum elastic 
coefficient $k_s$ is quite similar to the one for the thorax, and both are 
close to the estimation $25\,$kN/m, obtained in Section 
\ref{Adjustmentstothemodel}. Thus, the compressional force in Section 
\ref{Adjustmentstothemodel} is in agreement with the experimental data.\\

The forces $F_T$ and $F_D$ exhibited in Table~\ref{parametros_modelo_social} 
correspond to two different measurement conditions. The tolerance threshold 
$F_T$ represents the pain limit when pressure is applied during 1~s on the 
abdomen or sternum area. The death threshold represents the limit of fatality 
when pressure is applied during 15~s on the chest area. The total amount 
$15\,\times F_T$ is less than $F_D$ since pain occurs at an early 
stage before injury. We choose the $F_D$ value as a suitable estimation for 
the falling pressure in our model. This pressure is approximately 
$6227\,\mathrm{N}/0.068\,\mathrm{m}^2=91.6\,\mathrm{kN/m}^2$.  \\

The ratio between the torso surface and the body surface is 
\mbox{$S_T/S_B=0.039$}, according to Table~\ref{parametros_modelo_social}. This 
is the fraction of the human body where the pressure limit of
$91.6\,\mathrm{kN/m}^2$ is applied to (during 15~s). We will assume that this 
fraction remains approximately valid, regardless of the volume representation 
of the body. That is, for any chosen body model (\textit{i.e.} sphere, 
cylinder) enclosed by a surface $S$, we will assume that the 
piece of surface \mbox{$0.039\times S$} corresponds to an ``effective'' torso 
surface. The compressional pressure is supposed to be applied on this area. \\

In order to link the experimental data shown in 
Table~\ref{parametros_modelo_social} to the model parameters, we associate the 
pressure limit $91.6\,\mathrm{kN/m}^2$  to the ``effective'' compressional force 
defined in Section~\ref{Totaleffectiveforce}. That is, we postulate that 
the ``effective'' force limit for reaching unconsciousness or fatality is 
roughly

\begin{equation}
 f_e^\mathrm{max}=91.6\times 0.039\times S\label{experimental_eq}
\end{equation}
  
The surface $S$ needs to be specified for achieving a numerical value of 
$f_e^\mathrm{max}$. This value will be computed in 
Section~\ref{sec:Conditionfall}.\\

\section{\label{simulations}Numerical simulations}

\subsection{\label{numerical_geometry} Boundary and initial conditions}

We simulated the evacuation process of 225 pedestrians from a 20~m $\times$ 
20~m room with a single exit door. The door width was $L=1.2$~m, enough to 
allow up to two pedestrians to escape simultaneously. \\

\begin{figure*}[!htbp]
\centering
\subfloat[Initial state \label{torso}]{
\includegraphics[scale=0.45]{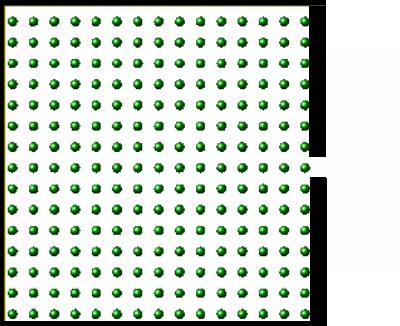} 
}\hfill
\subfloat[State at 10 seconds\label{proyeccion}]{
\includegraphics[scale=0.45]{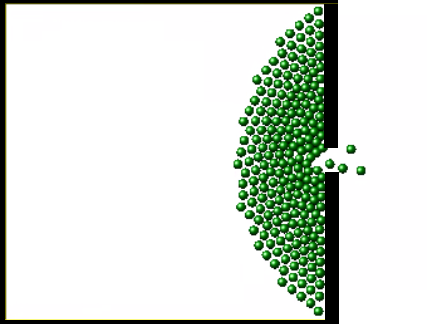}
}
\caption{\label{sim} Snapshots of an evacuation process from a 
20~m$\,\times\,$20~m room with a single door for 225 pedestrians. The picture 
on the left represents the initial configuration. The picture on the right 
represents the evacuation process after 10~s. All the individuals correspond to 
``moving pedestrians''. The black lines represents the walls. The desired 
velocity was $v_d$=6~m/s.} 
\end{figure*}

The process started with all the pedestrians inside the room and equally 
separated in a square arrangement, as shown in the Fig.~\ref{sim}. The 
occupancy density was set to 0.6~people/m$^2$, as suggested by healthy indoor 
environmental regulations \cite{mysen}. The pedestrians had random initial 
velocities computed from a Gaussian distribution (with null mean value). The 
rms value for the Gaussian distribution was close to 1~m/s. \\

The desired velocity $v_d$ was the same for all the individuals, meaning that 
all of them had the same anxiety level. At each time-step, however, the 
desired direction $\mathbf{e}_d$ was updated, in order to point to the exit. 
\\

The evacuation processes began with all the pedestrians moving randomly, but 
willing to go to the exit. If certain conditions were met (see 
Section~\ref{sec:Conditionfall}), any moving pedestrian could switch his 
(her) behaviour to the ``fallen'' pedestrian behaviour. Fallen pedestrians were 
those that remained at a fix position, that is, they did no longer move 
until the end of the evacuation process. \\

The evacuation processes were implemented on the {\sc Lammps} molecular 
dynamics simulator \cite{plimpton}. {\sc Lammps} was set to run on multiple 
processors. The chosen time integration scheme was the velocity Verlet 
algorithm with a time step of $10^{-4}\,$s. Any other parameter was the same 
as in previous works (see Refs.~\cite{frank2015,frank2011}). \\

We simulated between 30 and 360 processes for each evacuation  
situation (see figures caption for details). Data was recorded at time 
intervals of 0.05~s. The processes lasted for 300~s, but a few situations were 
also examined until 800~s (see Section~\ref{flux} for details). Only the 
evacuation processes shown in Section~\ref{pass_through} lasted until 100 
individuals were able to leave the room. \\

The explored anxiety levels ranged from relaxed situations ($v_d<$1~m/s) to 
highly stressing ones ($v_d$=8~m/s). Recall that the ``faster is slower'' 
effect occurs within this range.  \\

\subsection{Moving and fallen pedestrians}
\label{sec:Conditionfall}

We separated the pedestrians behavioural patterns into two categories: 
``moving'' individuals and ``fallen'' individuals. The former are those that 
move according to Eq.~(\ref{eq_mov}). The latter are those that are not able to 
move at all until the end of the evacuation process. Moving pedestrians, 
however, can switch to the fallen category, but fallen pedestrians always 
remain in that category. \\

The condition for a moving pedestrian to switch to a fallen pedestrian's 
behaviour is that the compressional pressure actuating on him (her) reaches the 
unconsciousness (or fatality) threshold for an uninterrupted time period of at 
least 15~s (see Section~\ref{experimentaldata}). This threshold is expressed by 
Eq.~(\ref{experimental_eq}). We simply modeled the pedestrian's body as spheres 
of radius $r_i=0.3\,$m (roughly, the neck-shoulder distance) and surface 
$S=4\pi r_i^2=1.13\,\mathrm{m}^2$. Therefore, the compressional threshold, in 
our model, became $f_e^\mathrm{max}=4030\,$N.\\

During the evacuation process simulation, we computed the ``effective'' 
force $f_e$ actuating on each moving pedestrian. This value was accumulated 
along time in a discrete variable $z_i$ as follows

\begin{equation}
 z_i=\left\{\begin{array}{ccl}
         z_i+1 & \mathrm{if} & f_e\geq f_e^\mathrm{max} \\
         z_i   &  \mathrm{if} & 0<f_e<f_e^\mathrm{max} \\
         0   &  \mathrm{if} & f_e=0 \\
        \end{array}\right.
\end{equation}

\noindent where $z_i$ was set to zero at the beginning of the process for each 
pedestrian $i$. Notice that the $z_i$ value is reseted whenever the 
``effective'' compresssional force $f_e$ vanishes, since the pedestrian's 
breathing restrictions are supposed to be released. The condition for the moving
pedestrian $i$ to become unconscious (\textit{i.e.} ``fallen'' pedestrian) is 
that $z_i=300$. Recall that the data recording was done every $0.05\,$s, and 
thus, the $z_i=300$ threshold represents a time period of 15~s since 
$300\times 0.05\,\mathrm{s}=15\,\mathrm{s}$.  \\ 

Any meeting situation between a moving pedestrian and a fallen one was handled 
in two possible ways: the moving pedestrian dodged the fallen pedestrian 
(similar to an obstacle avoidance), or, the moving pedestrian passed through 
the fallen one. The dodging scenario is examined from Section~\ref{flux} to 
Section~\ref{varyingvd}, while the passing-through scenario is examined in
Section~\ref{pass_through}. \\

In the dodging scenario, the forces actuating on the moving pedestrian due 
to the fallen individual were similar to the forces actuating between 
two neighboring moving pedestrians. That is, the moving pedestrian experienced 
the same repulsive feelings and sliding friction as if the fallen individual 
belonged to the ``moving'' category. In the passing-through scenario, neither 
repulsive feelings nor friction (due to the fallen individual) were present on 
the moving pedestrian. But, as explained in Section~\ref{passoverforce}, the 
desired force $\mathbf{f}_d$ was replaced by the ``passing-through'' force  
$\mathbf{f}_p$. The rest of the forces between moving pedestrians remained the 
same. \\

Recall that the ``passing-through'' scenario corresponds to a first approach on 
this kind of behavioral patterns. Therefore, we want to stress the fact that 
our model is as simple as we could imagine, in order to study the most basic 
effects out of the zero order approach of avoiding fallen pedestrians. \\

\section{\label{results}Results}

We divided our investigation into two different scenarios. From
Sections~\ref{flux} to \ref{varyingvd} we examined those situations where 
the moving pedestrians were only able to dodge the fallen individuals. In 
Section~\ref{pass_through} we relaxed this restriction, while allowing the 
moving pedestrians to pass-through the fallen individuals. A new \textit{working 
hypothesis} was introduced to achieve this behavioural pattern. Finally, in 
Section~\ref{pass_through} we compared the results obtained from these two main 
scenarios.  \\ 

\subsection{\label{flux}The outgoing flux at high pressure levels for the 
dodging scenario}

As a starting point, we examined the evacuees distribution for the first 300~s 
of the leaving process. All the individuals had a desired velocity of 6~m/s in 
order to achieve the ``faster is slower'' effect. This desired velocity 
also allowed many individuals to become unconscious (\textit{i.e.} fallen 
individuals).   \\

\begin{figure}
\centering
\includegraphics[scale=0.33]{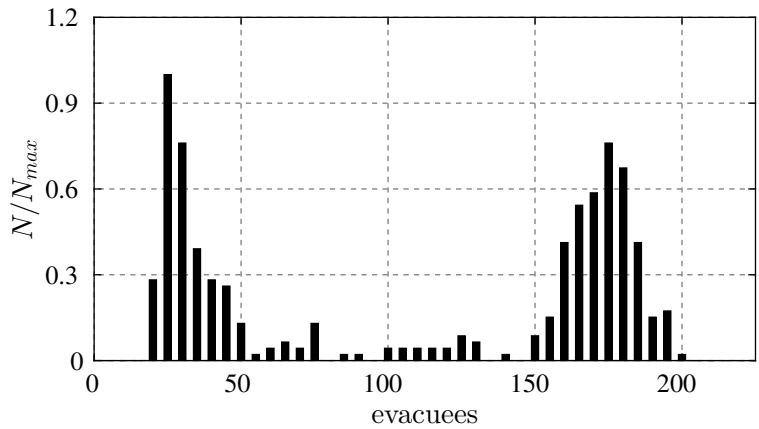}
\caption{\label{fig:3} Normalized distribution of evacuees along the first 
300~s. We considered bins of five pedestrians between the range of 
0 to 200 individuals. $N$ is the number of events corresponding to each bin. 
The plot is normalized with respect to the bin of maximum value $N_{max}$. 
Data was recorded from 360 evacuation processes. The desired velocity was 
$v_d=6\,$m/s.}
\end{figure}

Fig.~\ref{fig:3} shows the corresponding histogram for the evacuees distribution 
(see caption for details). As can be seen, the most probable number of 
evacuees lies within two separated intervals: the lower interval ranging 
between 0 and 50 evacuees, and the upper interval ranging between 150 and 200 
evacuees. The intermediate interval in between, say, 50 to 150 evacuees is 
very unlikely.  \\

The drawback for the evacuees distribution in Fig.~\ref{fig:3} is that we get no 
information on whether the evacuation processes fail for the lower and 
intermediate intervals (that is, the exit gets blocked), or, if only a 
``slowing down'' is present in any of these processes. The latter means that the 
time delays between two outgoing individuals are very large, and thus, the 
evacuation does not finish by 300~s. \\    

\begin{figure}
\centering
\includegraphics[scale=0.3]{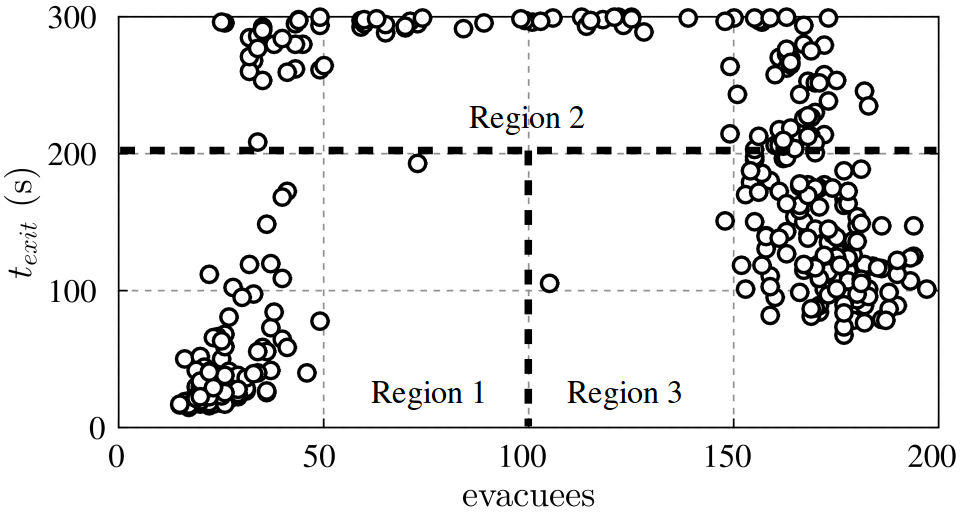}
\caption{\label{fig:4} Outgoing time $t_{exit}$ of the last evacuee (before 
300~s) as a function of  the number of evacuees until 300~s. Each circle 
represents an evacuation process. 360 processes are exhibited in the plot. 
The dashed lines represents a qualitative boundary for three regions, labeled 
as ``region 1'', ``region 2'' and ``region 3''. The desired velocity 
was $v_d=6\,$m/s.}
\end{figure}

In order to get a better understanding of the evacuation process, we recorded 
the time when the last pedestrian left the room before 300~s for each 
process. We also did this recording at 800~s (not shown) as a 
crosscheck.  Fig.~\ref{fig:4} shows the recorded time $t_{exit}$ vs. the number 
of pedestrians that left the room for the 300~s case. \\

The data points in Fig.~\ref{fig:4} show that the number of leaving 
pedestrians mostly lie in the lower and upper intervals, as exhibited in 
Fig.~\ref{fig:3}. But, the leaving time for the last pedestrian 
$t_{exit}$  (before 300~s) does usually not exceed the 200~s. We can 
envisage some kind of correlation between the leaving time for the last 
individual (before 300~s) and the total number of evacuees. \\  

Notice in Fig.~\ref{fig:4} that $t_{exit}$ is close to 300~s for the number of 
leaving pedestrians that lie between 50 and 150 individuals. The crosscheck 
with the recording at 800~s (not exhibited) shows that many of the data points 
that formerly lied in the 50 to 150 evacuee interval (see Fig.~\ref{fig:4}), 
actually move to the upper interval (beyond 150 evacuees). This means that the 
50 to 150 evacuee interval corresponds to not completely finished  processes, 
and consequently, to a ``slowing down'' in the evacuation.  \\ 

Fig.~\ref{fig:4} summarizes three qualitative situations. These are roughly 
separated by the dashed lines. The first situation (labeled 
as ``region 1'') corresponds to those processes where a small fraction of the 
pedestrians are able to leave the room and the evacuation virtually ceases 
after 200~s. The second situation (labeled as ``region 2'') corresponds to a 
``slowing down'' in the evacuation process, with an uncertain ending time. The 
third situation (labeled as ``region 3'') actually finished when almost all of 
the individuals (beyond 150) left the room.  \\ 

The existence of three qualitatively different situations is a novel result. 
The fallen people seem to affect the evacuation performance in different ways 
if the surrounding pedestrians are only able to avoid them. Thus, we decided to 
deep into the fallen people behaviour for the further understanding of this 
new dynamic in the high pressure scenario ($v_d=6\,$m/s). \\

\begin{figure}
\centering
\includegraphics[scale=0.3]{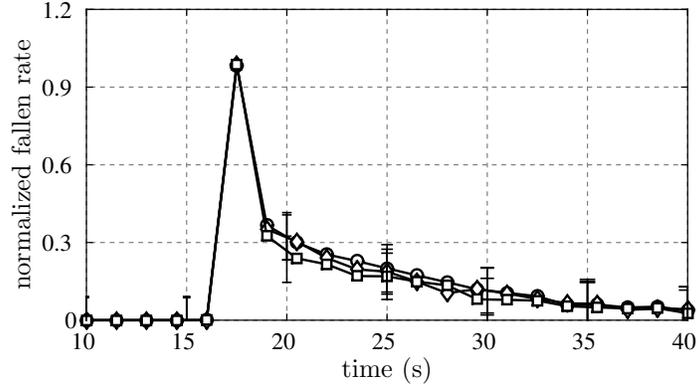}
\caption{\label{fallen_per_time} Normalized fallen rate vs. time (in seconds).  
$\bigcirc$ represent the ``region 1'' processes. \mbox{$\diamondsuit$ 
represent} the ``region 2'' processes. \mbox{$\Box$ represent} the ``region 
3'' processes. Mean values were computed from (approx.) 120 realizations. The 
curves are normalized to have its maximum at unity. The error bars corresponds 
to $\pm\sigma$ (one standard deviation). The desired velocity was 
$v_d=6\,$m/s.} 
\end{figure}

Fig.~\ref{fallen_per_time} shows the (mean) number of fallen pedestrians per 
unit time along the evacuation process for each qualitative region. Clearly, 
the maximum rate of fallen pedestrians occurs at the beginning of the 
evacuation process, say, at approximately 20~s. This is in 
agreement with the fact that high pressures should be present for
least for 15~s before the individual becomes unconscious (see 
Section~\ref{experimentaldata}). Furthermore, Fig.~\ref{fallen_per_time} 
indicate that the pressure in the bulk surmounts the injury limit from the very 
beginning of the evacuation process.  However, after the first 20~s, the rate of 
fallen individuals slows down, regardless of the locus where the process lies 
in Fig.~\ref{fig:4}. The qualitative different situations corresponding to 
the locus in Fig.~\ref{fig:4} does not actually depend on the rate of 
unconsciousness. We also checked over that these situations do not depend on 
the total number of fallen individuals (roughly, 20 individuals in our 
simulations).   \\

\begin{figure}
\centering
\includegraphics[scale=0.3]{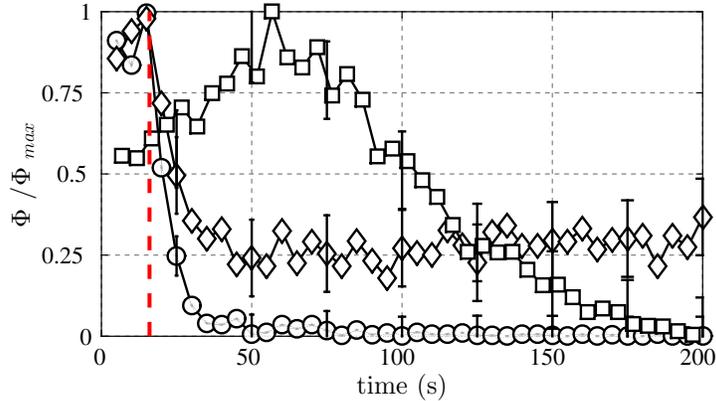}
\caption{\label{fig:5} Normalized evacuees flow rate ($\Phi 
/\Phi_{\textup{max}}$)  vs. time. $\bigcirc$ represent the ``region 1'' 
processes. \mbox{$\diamondsuit$ represent} the ``region 2'' processes. 
\mbox{$\Box$ represent} the ``region 3'' processes. The vertical red dashed 
line corresponds  to the time stamp of maximum number of fallen pedestrians per 
unit time (see Fig.~\ref{fallen_per_time}). Mean values were computed from 
(approx.) 120 realizations. The curves are normalized to have its maximum at 
unity. The error bars corresponds to $\pm\sigma$ (one standard deviation). The 
desired velocity was $v_d=6\,$m/s.}
\end{figure}

Besides the rate of fallen pedestrians, we analyzed the  flow rate of surviving 
individuals. Fig.~\ref{fig:5} exhibits the flow rate of the leaving pedestrians 
along time. Three data sets are shown, each one representing the mean flow 
value for each corresponding region in Fig.~\ref{fig:4}. A vertical dashed line 
(red in the on-line version) also represents the time when the maximum rate of 
fallen people occurs. Notice that the (mean) flow rate for the ``region 3'' 
processes is qualitatively different from the ones corresponding to the 
``region 1'' and the ``region 2'' processes. The former has a positive slope 
until 50~s, while regions 1-2 have negative or null slopes. Therefore, the 
``region 3'' processes manage to keep a high rate of people leaving the room, 
while the other two situations makes it harder (or even impossible) for the 
individuals to escape.  
\\

Fig.~\ref{fig:5} also gives us a better understanding of Fig.~\ref{fig:4} for 
the different behaviours between the ``region 1'' and ``region 2'' processes. 
Both situations have a diminishing flow rate, according to Fig.~\ref{fig:5}. 
But, the  flow rate of the ``region 2'' processes is non-vanishing, although 
weak. The number of evacuees above 50 in ``region 2'' (see Fig.~\ref{fig:4}) 
can be explained by the weak flow rate beyond 50~s appearing in 
Fig.~\ref{fig:5}. This is in agreement with the mentioned uncertainty in the 
ending time for the ``region 2'' processes. On the contrary, the vanishing flow 
rate for the ``region 1'' processes means that the evacuation process finishes 
after a relatively short time period. Consequently, the expected $t_{exit}$ is 
relatively low, as shown in  Fig.~\ref{fig:4}. \\  

A few conclusions can be outlined from the above analysis. For the dodging 
scenario (and high pressures) three situations appear to be possible. The first 
situation (\textit{i.e.} ``region 1'' processes) occurs when the evacuation 
ceases in a short time period and a small fraction of the pedestrians are able 
to leave the room. The second situation (\textit{i.e.} ``region 2'' processes) 
has a very similar performance as the first situation at the beginning of the 
process, but instead of ceasing after this time period, it ``slows down'', 
like a ``leaking'' process. The ``slow down'' seems endless because the 
``leaking'' delays the evacuation to very long time periods. The third 
situation (\textit{i.e.} ``region 3'' processes) corresponds to high flow 
rates, allowing a large fraction of the pedestrians to leave the room.   \\

Surprisingly, the rate of fallen pedestrians is actually not relevant for 
allowing one of the three situations. Thus, the spatial distribution of the 
fallen individuals should be analyzed next.\\

\subsection{\label{pressure} The social pressure for the dodging scenario}

In Section~\ref{flux} we analyzed the rate of fallen individuals and the flow 
rate of the outgoing pedestrians. As a second step in the investigation, we 
studied the pressure patterns inside the bulk. We first computed the bulk mean 
social pressure due to the moving pedestrians (that is, excluding the fallen 
individuals), and then, we studied the spatial distribution of the pressure 
for all the individuals (including the fallen pedestrians).\\

\begin{figure*}
\subfloat[\label{pres_flujo_z1}]{
\includegraphics[scale=0.4]{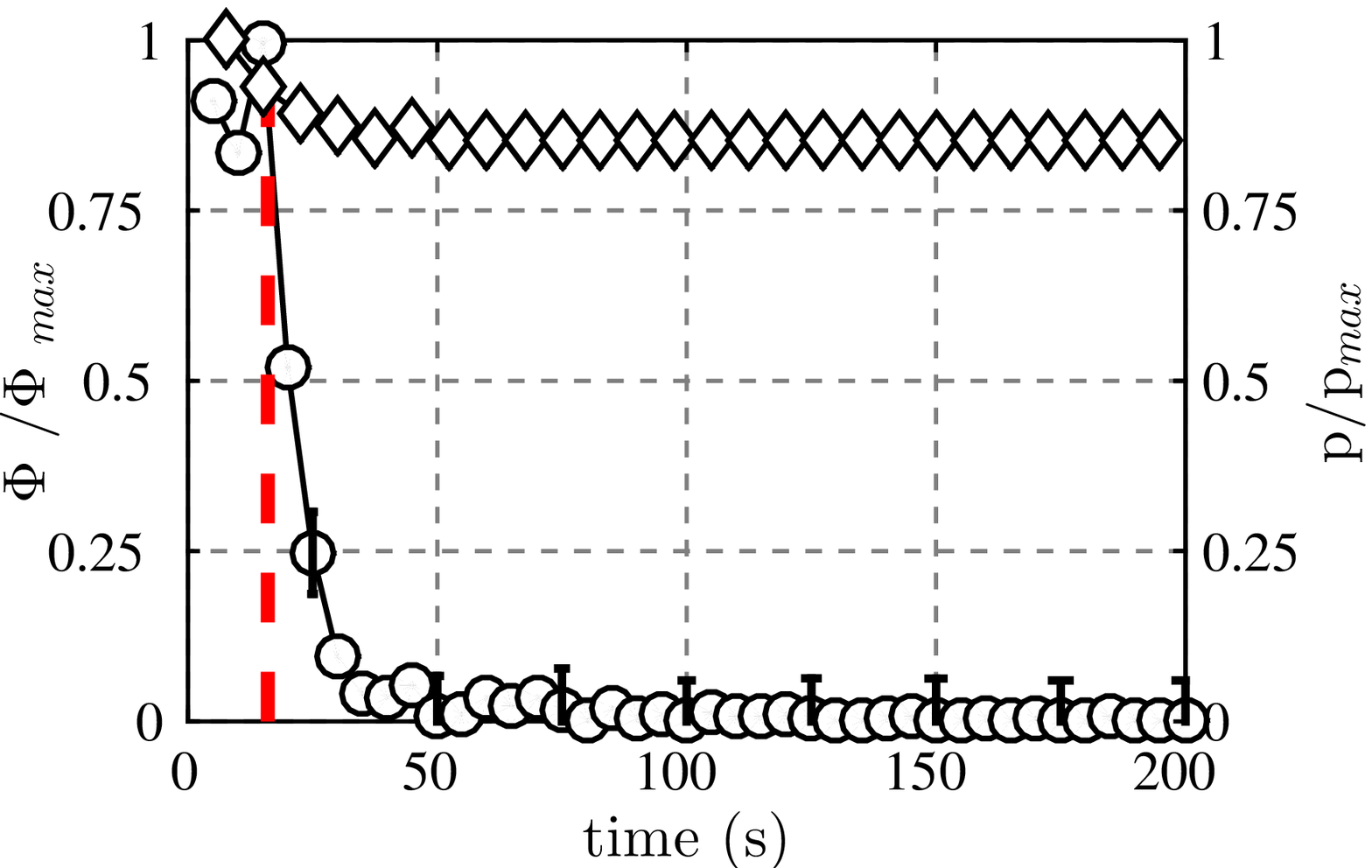}
}\hfill
\subfloat[\label{pres_evac_z1}]{
\includegraphics[scale=0.4]{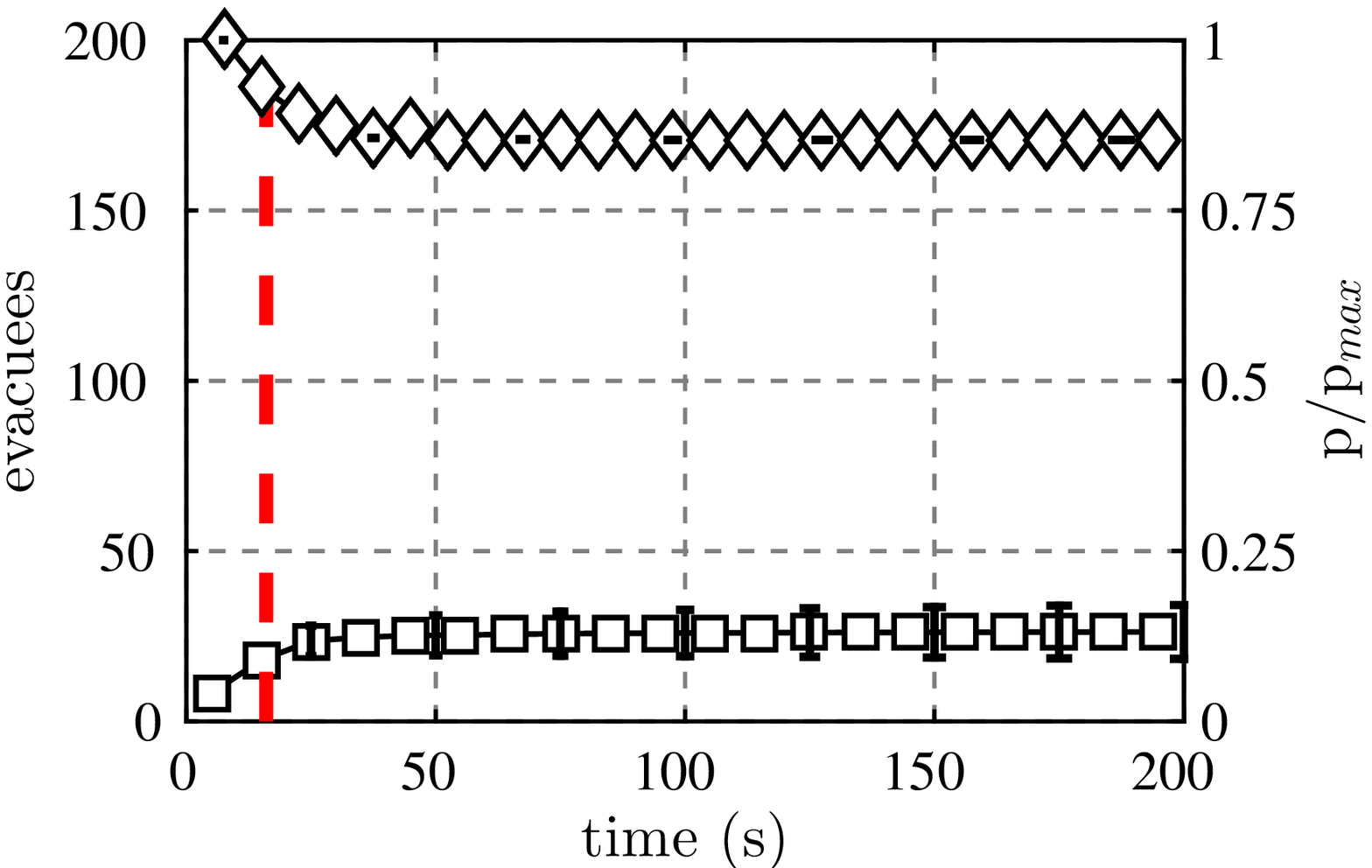}
}
\hfill
\subfloat[\label{pres_flujo_z3}]{
\includegraphics[scale=0.4]{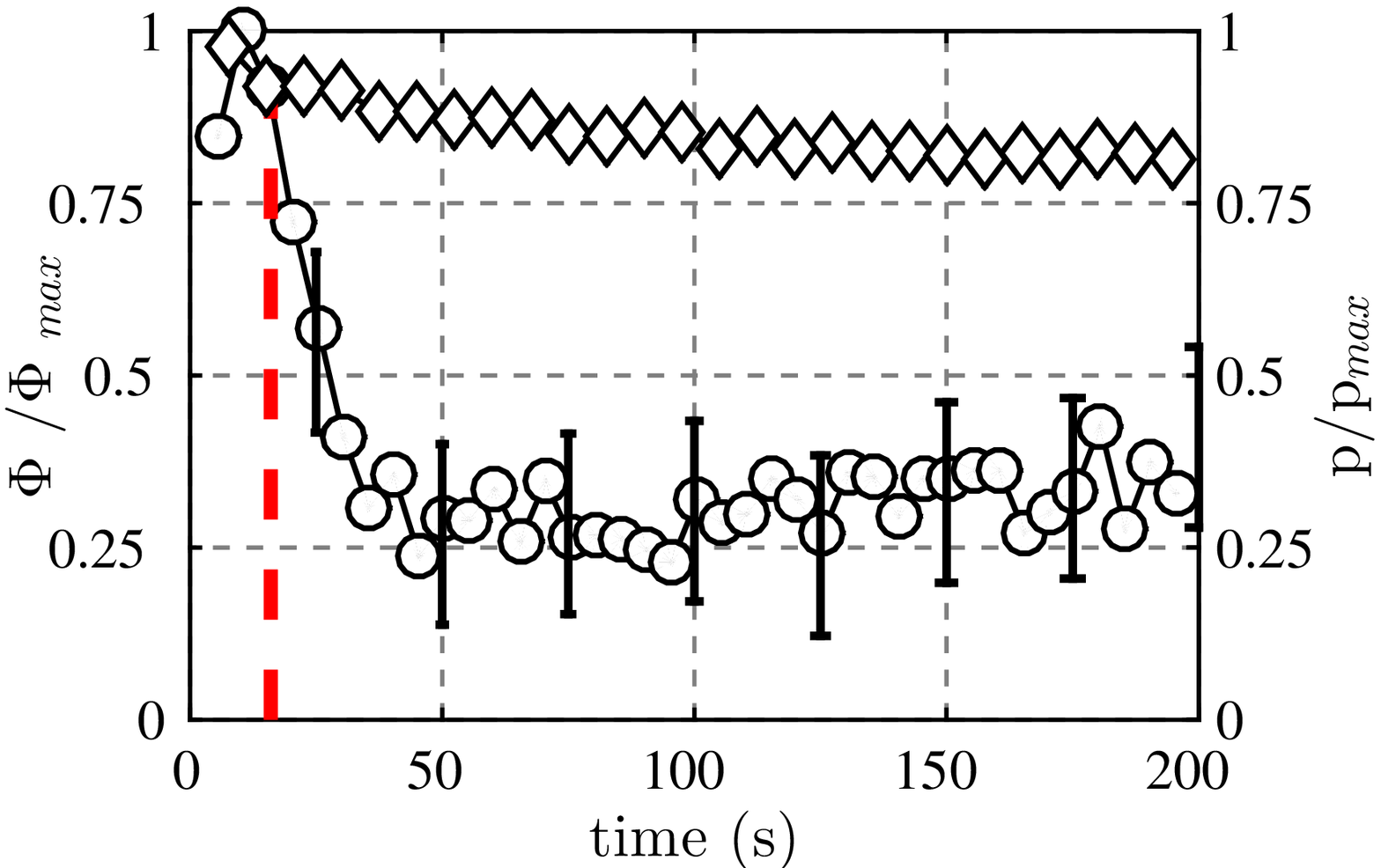}
}
\hfill
\subfloat[\label{pres_evac_z3}]{
\includegraphics[scale=0.4]{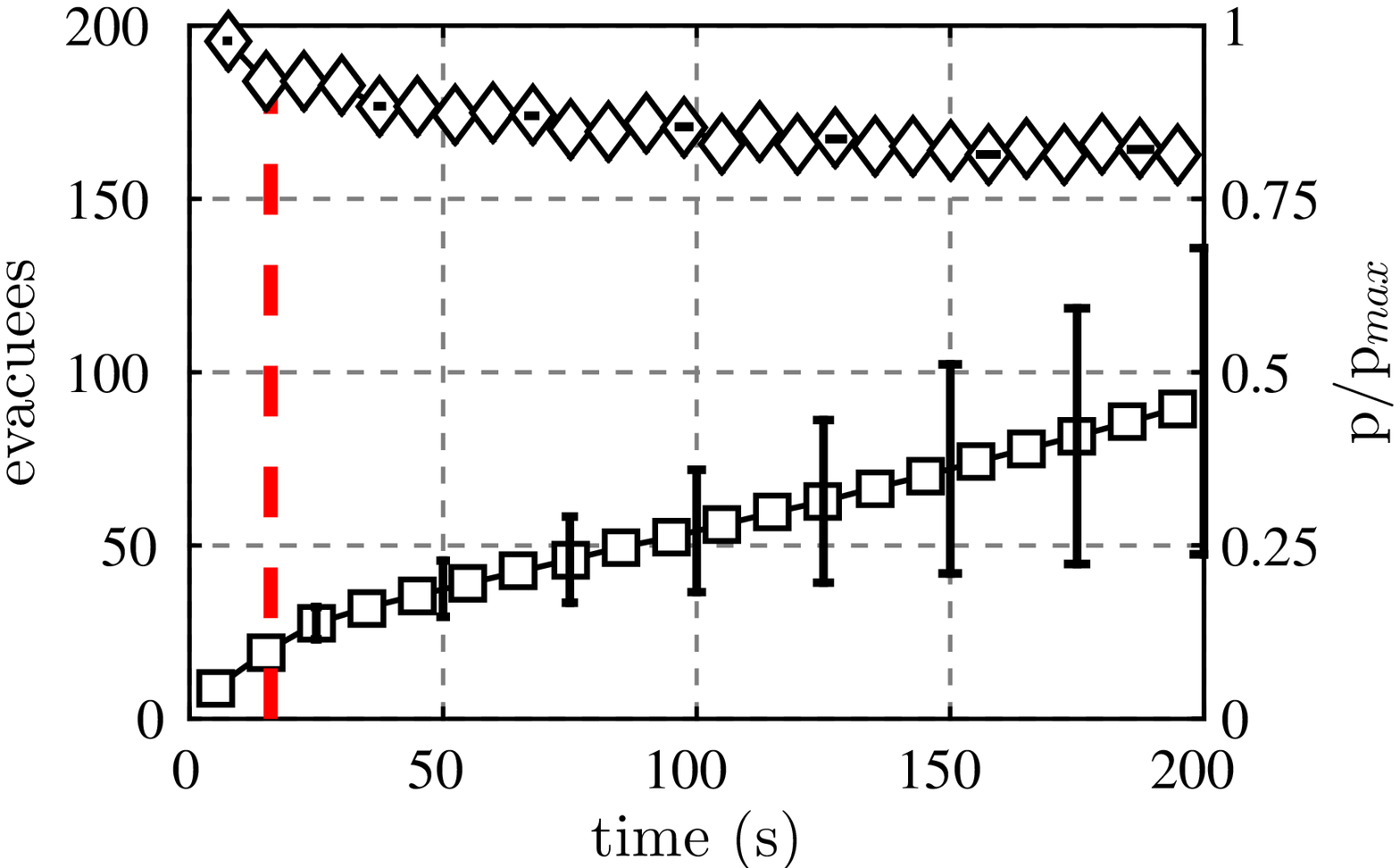}
}
\hfill
\subfloat[\label{pres_flujo_z4}]{
\includegraphics[scale=0.4]{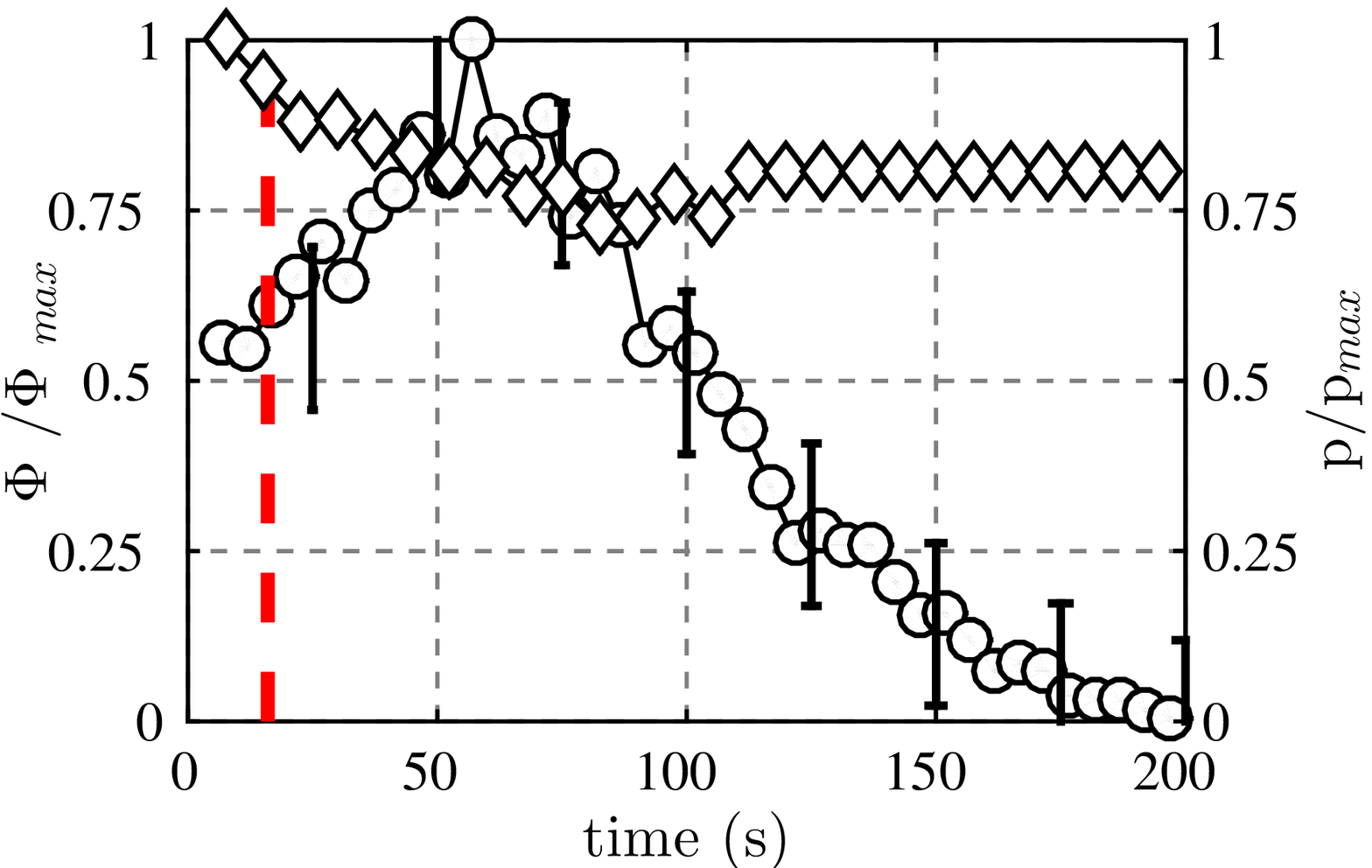}
}
\hfill
\subfloat[\label{pres_evac_z4}]{
\includegraphics[scale=0.4]{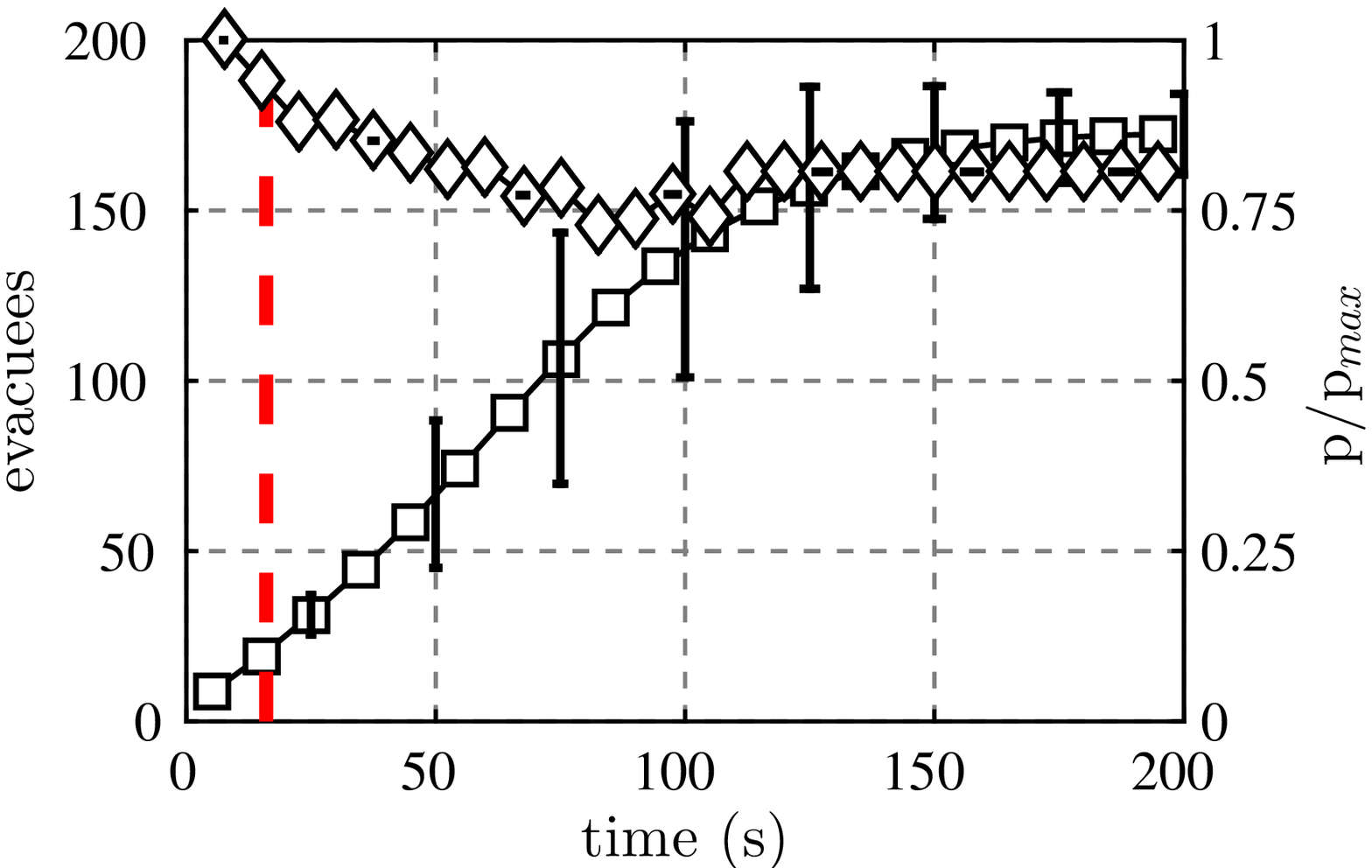}
}
\caption{\label{fig:presiones_flujos_evacuados} Normalized flow rates, number 
of evacuees and social pressures along time. For (a), (c) and (e) 
plots:~$\bigcirc$ represent the normalized evacuees flow rate ($\Phi 
/\Phi_{\textup{max}}$) and $\diamondsuit$ represent the normalized mean social 
pressure (p/p$_{\textup{max}}$) vs. time. For (b), (d) and (f) plots:~$\Box$ 
represent the amount of evacuees and $\diamondsuit$ represent the normalized 
mean social pressure (p/p$_{\textup{max}}$) vs. time. The (a) and (b) 
plots correspond to ``region 1''. The (c) and (d) plots correspond to ``region 
2''. The (e) and (f) plots correspond to ``region 3''. Only moving pedestrians 
contributed to the mean social pressure. The vertical red dashed line 
corresponds to the time stamp for maximum number of fallen pedestrians per 
unit time (see Fig.~\ref{fallen_per_time}). Mean values were computed from 
(approx.) 120 realizations. The curves are normalized to have its maximum at 
unity. The error bars corresponds to $\pm\sigma$ (one standard deviation). The 
desired velocity was $v_d=6\,$m/s.} 
\end{figure*}

Fig.~\ref{fig:presiones_flujos_evacuados} shows the mean social pressure of the 
moving pedestrians in the bulk along the evacuation process. The leaving flow 
rate and the total number of evacuees are also included for comparison 
reasons. The data sets are separated into the qualitative situations described 
in Section~\ref{flux} (\textit{i.e} regions 1-3). \\

The mean social bulk pressure shown in 
Fig.~\ref{fig:presiones_flujos_evacuados} converges approximately to a fixed 
fraction of the maximum pressure value, regardless of the flow rate pattern and 
the amount of evacuees. This means that some individuals remain in the room at 
the end of the process, whatever the locus of this process in Fig.~\ref{fig:4}. 
We checked over this result by running some process animations (not shown) and 
we found that this asymptotic pressure corresponds to the pressure actuating on 
those individuals that actually get locked by fallen pedestrians. 
 \\

We can see in Figs.~\ref{pres_flujo_z1} and 
\ref{pres_flujo_z3} (and the corresponding Fig.~\ref{pres_evac_z1}, 
\ref{pres_evac_z3}) that the pressure slope is always negative (or 
vanishing). However, it can be noticed that while the pressure settles 
at the same time as the flow rate in Fig.~\ref{pres_flujo_z1}, it does not in 
Fig.~\ref{pres_flujo_z3}. That is, the ``region 2'' processes represented in 
Fig.~\ref{pres_flujo_z3} continue loosing pressure during the constant flow 
rate interval (say, beyond 50~s). We can envisage here some kind of 
``slowly controlled'' evacuation situation that is not present in the ``region 
1'' processes. This means that the pressure reduction and the leaving flux 
are related in some way. \\  

The pressure slope shown in Fig.~\ref{pres_flujo_z4} and 
Fig.~\ref{pres_evac_z4} for the ``region 3'' processes is negative along the 
first 75~s (approximately). It slightly changes sign near 100~s, and 
immediately after, vanishes. Notice that this pressure pattern is qualitatively 
different from the ones shown in Figs.~\ref{pres_flujo_z1} and 
\ref{pres_flujo_z3}. Here, the pressure lose occurs while the flow rate 
increases or remains almost unchanged (for a short period of time). But, while 
the pressure slope changes sign and vanishes, the flow rate decreases (see 
Fig.~\ref{pres_flujo_z4} beyond 100~s). We can not envisage a 
``slowly controlled'' evacuation situation, as for the ``region 2'' situation. 
This suggests that the pedestrians might be leaving the room in a somehow 
different way. \\

\begin{figure*}
\subfloat[Region 1\label{press_zone1}]{
\includegraphics[scale=0.9]{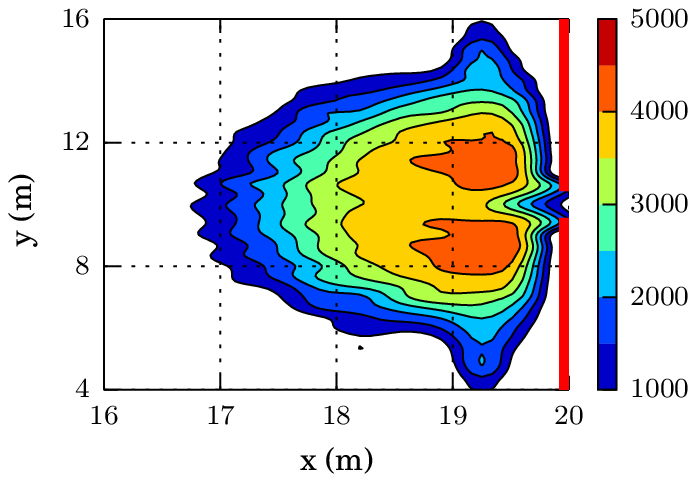}
}\hfill
\subfloat[Region 2\label{press_zone3}]{
\includegraphics[scale=0.9]{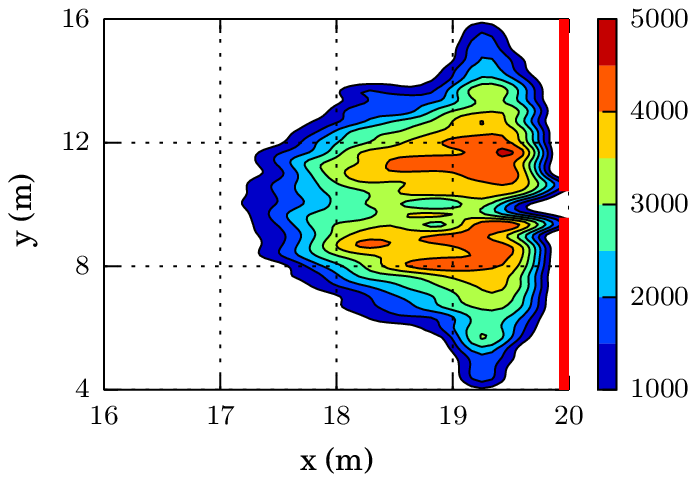}
}
\hspace{-6mm} \subfloat[Region 3\label{press_zone4}]{
\includegraphics[scale=0.9]{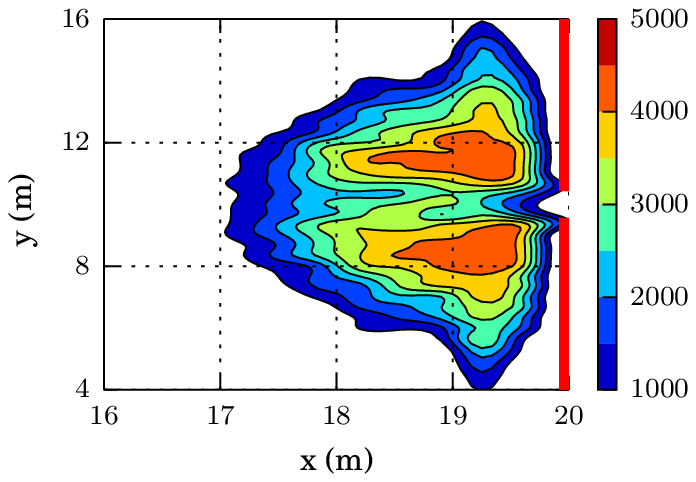}
}
\hfill
\subfloat[No fallen pedestrians\label{press_no_deaths}]{
\includegraphics[scale=0.9]{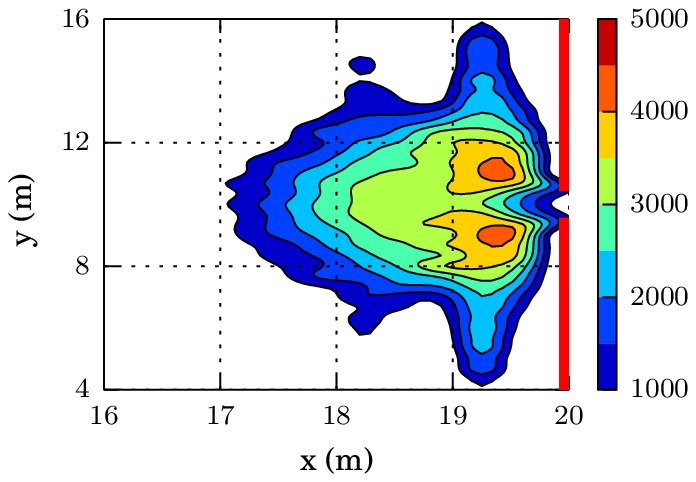}
}
\hfill
\caption{(Color on-line only) Mean social pressure contour lines computed from 
30 evacuation processes for the first 100~s. Fig.~\ref{press_no_deaths} 
represent corresponds to evacuation processes with non fallen (unconscious) 
individuals. The scale bar on the right is expressed in 
N.m$^{-1}$ units (see text for details). We included the social pressure over 
both types of pedestrians (moving and fallen). The red lines at $x~=$~20~m 
represent the walls on the right of the room. The pedestrian's desired velocity 
was $v_d=6\,$m/s. The contour lines were computed on a square grid of 
$1\,\mathrm{m}\times1\,\mathrm{m}$ and then splined to get smooth curves. Level 
colors can be seen in the on-line version only. 
\label{fig:campo_presiones}
} 
\end{figure*}

We next examined the mean social pressure contour maps for the three main loci 
represented in Fig.~\ref{fig:4}. The contour maps are shown in 
Fig.~\ref{fig:campo_presiones}. It also includes the situation where the 
pedestrians are not allowed to become unconscious (see 
Fig.~\ref{press_no_deaths}). Notice that the contour maps include the pressure 
actuating on \textit{all} the crowd (that is, either the conscious and 
unconscious pedestrians).  \\

The higher pressure zones in all the contour maps appear on the sides of the 
exit (see Fig.~\ref{fig:campo_presiones}). But the ``region 1'' 
processes (Fig.~\ref{press_zone1}) exhibit a qualitative difference in the 
middle of the room with respect to the other processes (Fig.~\ref{press_zone3} 
and Fig.~\ref{press_zone4}).  The former shows a widely 
spread high pressure area centered at the mid-path $y=10\,$m. Instead, 
Fig.~\ref{press_zone3} and Fig.~\ref{press_zone4} show  a 
low pressure path along $y=10\,$m. \\

Recall that the flow rate vanishes for the ``region 1'' situation. Thus, 
Fig.~\ref{press_zone1} represents the pressure map when the pedestrians are 
not able to leave the room. Notice that this pressure map is opposed to 
the one shown in Fig.~\ref{press_no_deaths} where (all) non-unconscious 
pedestrians can manage to get out. This outgoing flow diminishes the bulk 
pressure, specially in the middle of the room (see Fig.~\ref{press_no_deaths}). 
See further details in Section~\ref{pass_through}.  \\

The low pressure mid-path appearing in Fig.~\ref{press_zone3} and 
Fig.~\ref{press_zone4} can be associated to the non-vanishing pedestrian flow 
rates for ``region 2'' and ``region 3'', respectively. However, the ``region 
3'' mid-path pattern resembles better the one with no unconscious pedestrians 
(see Fig.~\ref{press_no_deaths}). This fact suggests that the mid-path 
configuration is responsible for the performance differences between the 
``region 2'' and the ``region 3'' situations.   \\

The possible evacuation situations for the dodging scenario can be summarized
as follows. The first situation (\textit{i.e.} ``region 1'' processes) 
only allows the evacuation for a short time period after the rate of fallen 
pedestrians reaches a maximum. No low pressure paths remain open after the 
evacuation becomes frustrated. The low pressure paths only remain open (during 
long time periods) for the second and third situation (\textit{i.e.} ``region 
2'' and ``region 3'', respectively). However, some connection appears to 
exist between the evacuation performance of each situation and the path 
configuration. The third situation resembles better the evacuation processes 
with no unconscious pedestrians.  \\

\subsection{\label{morphology}The evacuation pathway for the dodging 
scenario}

\begin{figure*}
\subfloat[Total blocking (region 1) \label{fig:7}]{
\includegraphics[scale=0.46]{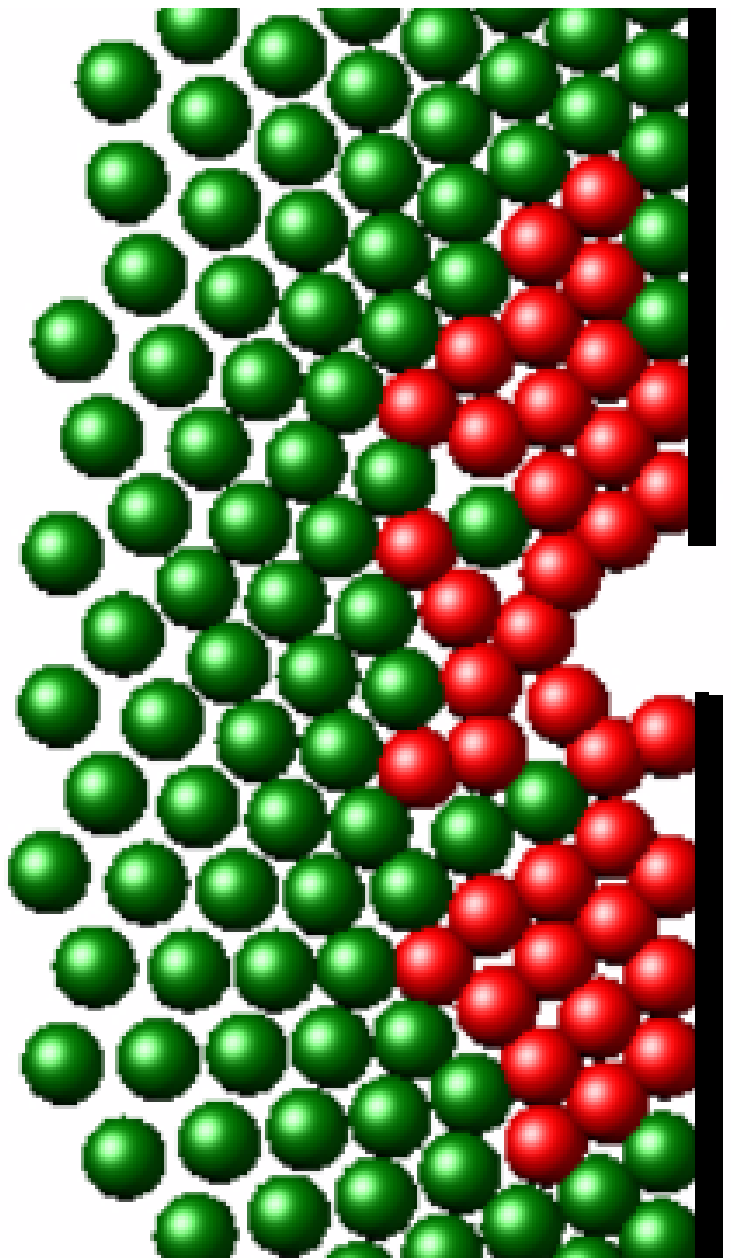}
}\hfill
\subfloat[Partial blocking (region 1) \label{fig:8}]{
\includegraphics[scale=0.46]{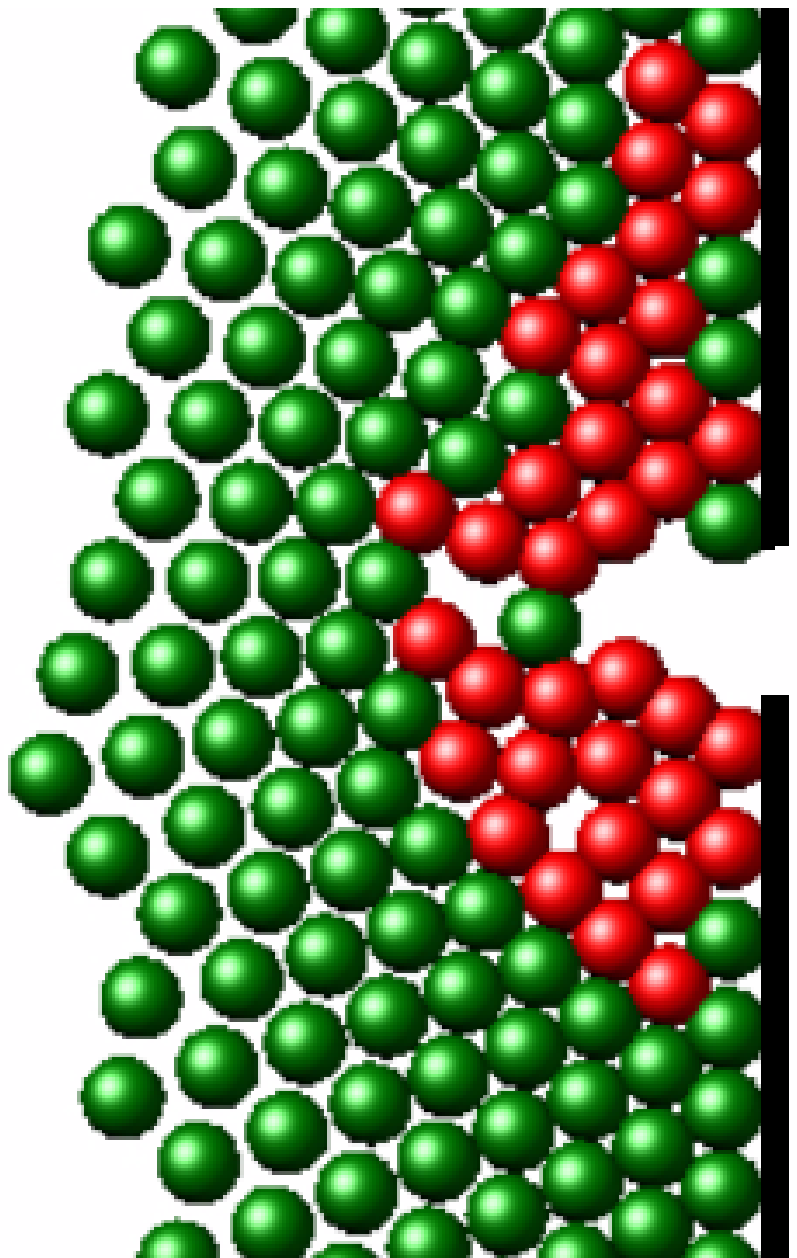}
}
\hfill
\subfloat[Narrow pathway (region 2) \label{fig:9}]{
\includegraphics[scale=0.46]{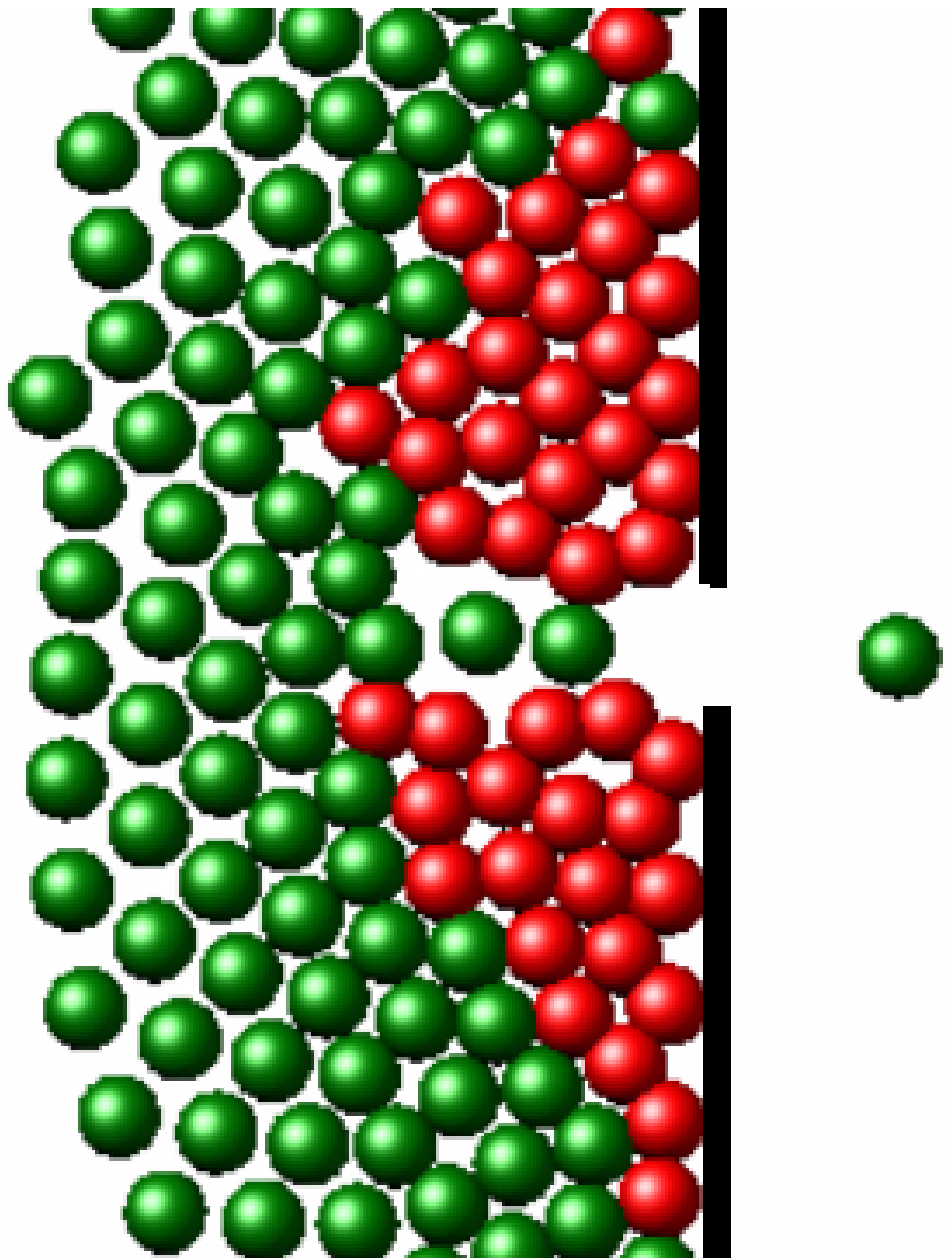}
}
\hfill
\subfloat[Wide pathway (region 3) \label{fig:10}]{
\includegraphics[scale=0.46]{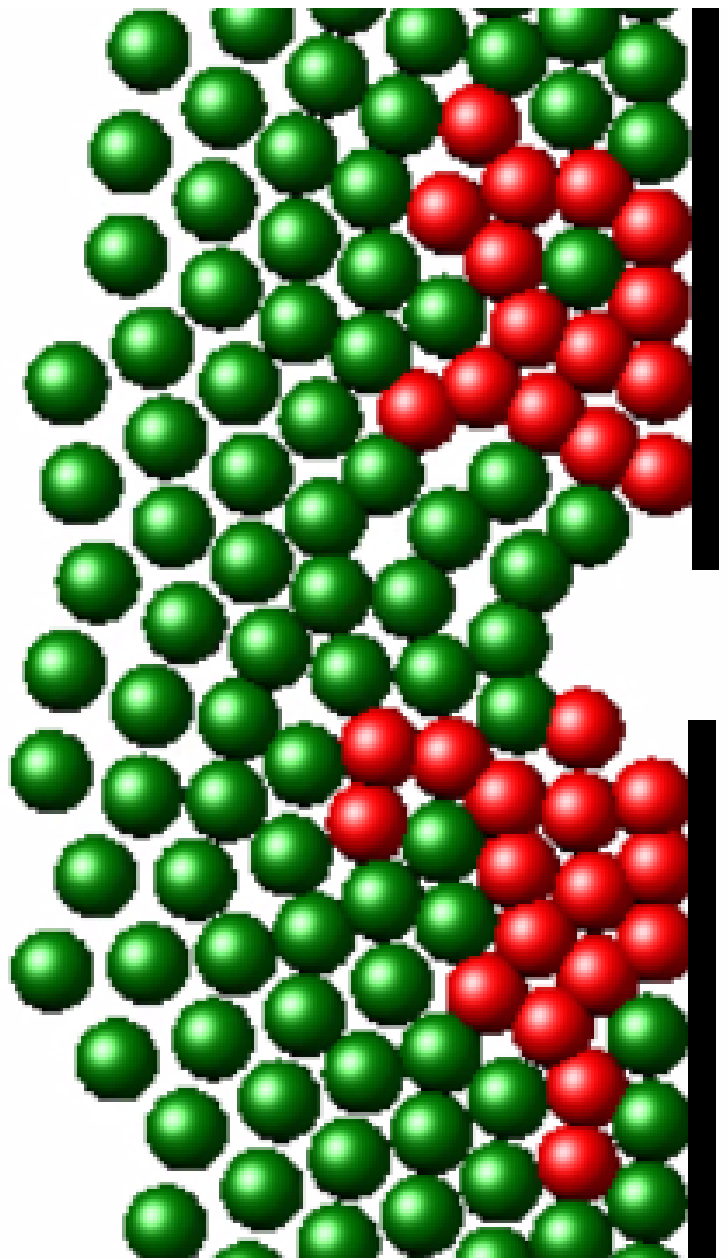}
}

\caption{\label{fig:types_evacuation}(Color on-line only) Snapshots of different 
evacuation processes for each region in the first 100~s of the processes. 
Moving and fallen pedestrians are represents in green and red circles, 
respectively. The black lines represent the walls on the right of the room. The 
pedestrian's desired velocity was $v_d=6\,$m/s.} 
\end{figure*}

Recall once again the process loci shown in Fig.~\ref{fig:4}. We 
examined separately the process animations for the three situations labeled as 
regions 1, 2 and 3. Fig.~\ref{fig:types_evacuation} shows four 
representative snapshots for these evacuation processes. The snapshots 
were recorded at 100~s, that is, at the stage where each situation can be 
differentiated easily. \\

Fig.~\ref{fig:7} and Fig.~\ref{fig:8} correspond to two representative 
snapshots for the ``region 1'' situation. The unconscious (fallen) pedestrians 
are blocking the exit in both pictures. However, only one blocking cluster 
appears in Fig.~\ref{fig:7}, while two blocking clusters can be seen in 
Fig.~\ref{fig:8}. These correspond to a \textit{total} blocking situation and a
\textit{partial} blocking situation, respectively, as defined in 
Section~\ref{human}. No pathway exists at all for the moving pedestrians to 
leave the room, in agreement with the corresponding (mean) flow rate shown in
Fig.~\ref{pres_flujo_z1} and the contour map shown in Fig.~\ref{press_zone1}. 
 \\

Fig.~\ref{fig:9} and Fig.~\ref{fig:10} correspond to two representative 
snapshots for the ``region 2'' and ``region 3'' situations, respectively. Both 
situations exhibit an available pathway for the moving pedestrians to leave the 
room. This means that the outgoing flow is non-vanishing, as reported in 
Fig.~\ref{pres_flujo_z3} and Fig.~\ref{pres_flujo_z4}. Thus, the snapshots 
confirm that a mid-path configuration is actually responsible for allowing the 
individuals to leave the room. The difference between Fig.~\ref{fig:9} and 
Fig.~\ref{fig:10} corresponds, however, to the pathway width. This path is 
approximately one pedestrian width (0.6~m) for the ``region 2'' situation, 
while it appears wider for the ``region 3'' situation. The contour maps in
Fig.~\ref{press_zone3} and Fig.~\ref{press_zone4} resemble quite accurately 
this difference.\\  

It is immediate that the wider the leaving pathway, the better evacuation 
performance. But a close examination of the animations for the ``region 2'' 
situation shows that the pedestrians leave the room intermittently, following a 
stop-and-go behaviour. This is qualitatively different from the ``region 3'' 
situation, where more than one individual can leave the room almost 
simultaneously. The stop-and-go behaviour is responsible for the regular flow 
in Fig.~\ref{pres_flujo_z3}, resembling a ``leaking-like'' process. On the 
contrary, the ``region 3'' situation allows an increasing number of 
pedestrians to leave the room (see Fig.~\ref{pres_flujo_z4}), until no more  
unlocked pedestrians are available in the room. \\ 

The above analysis from the snapshots and animations (not shown) summarizes as 
follows. The pedestrians located on the sides of the door experience the higher 
pressure in the bulk, and thus, have the higher probability to become 
unconscious. The unconscious (fallen) pedestrians may or may not block the exit. 
If a partial or total blocking occurs (as defined in 
Section~\ref{human}), the outgoing flow immediately vanishes and the evacuation 
becomes frustrated. The ``region 1'' resembles this situation. \\

If the unconscious (fallen) pedestrians do not block the exit, the remaining 
pathway plays an important role. For narrow pathways (\textit{i.e.} width close 
to 0.6~m), the overall evacuation slows down due to a stop-and-go dynamic along 
the pathway. For wider pathways, the individuals can manage to get out easily, 
and consequently, the evacuation process improves. \\   

Notice that all the pedestrians are located in a square arrangement at the 
beginning of the process. The initial velocities are set to random (according 
to a Gaussian distribution). Thus, as a first thought, we might expect the 
three studied situations to be equally likely. In Section~\ref{varyingvd} we 
will show that this is not 
actually the case. \\

\subsection{\label{varyingvd}The role of the desired velocity for the dodging 
scenario}

We examined the dodging scenario for the desired velocity $v_d=6\,$m/s 
through Section~\ref{flux} to Section~\ref{morphology}. We now vary 
the desired velocity from 4.5~m/s to 8~m/s. For desired velocities below 
4.5~m/s, no pedestrians become unconscious. \\

\begin{figure*}
\subfloat[ \label{fig:17}]{
\hspace{-4mm} \includegraphics[scale=0.4]{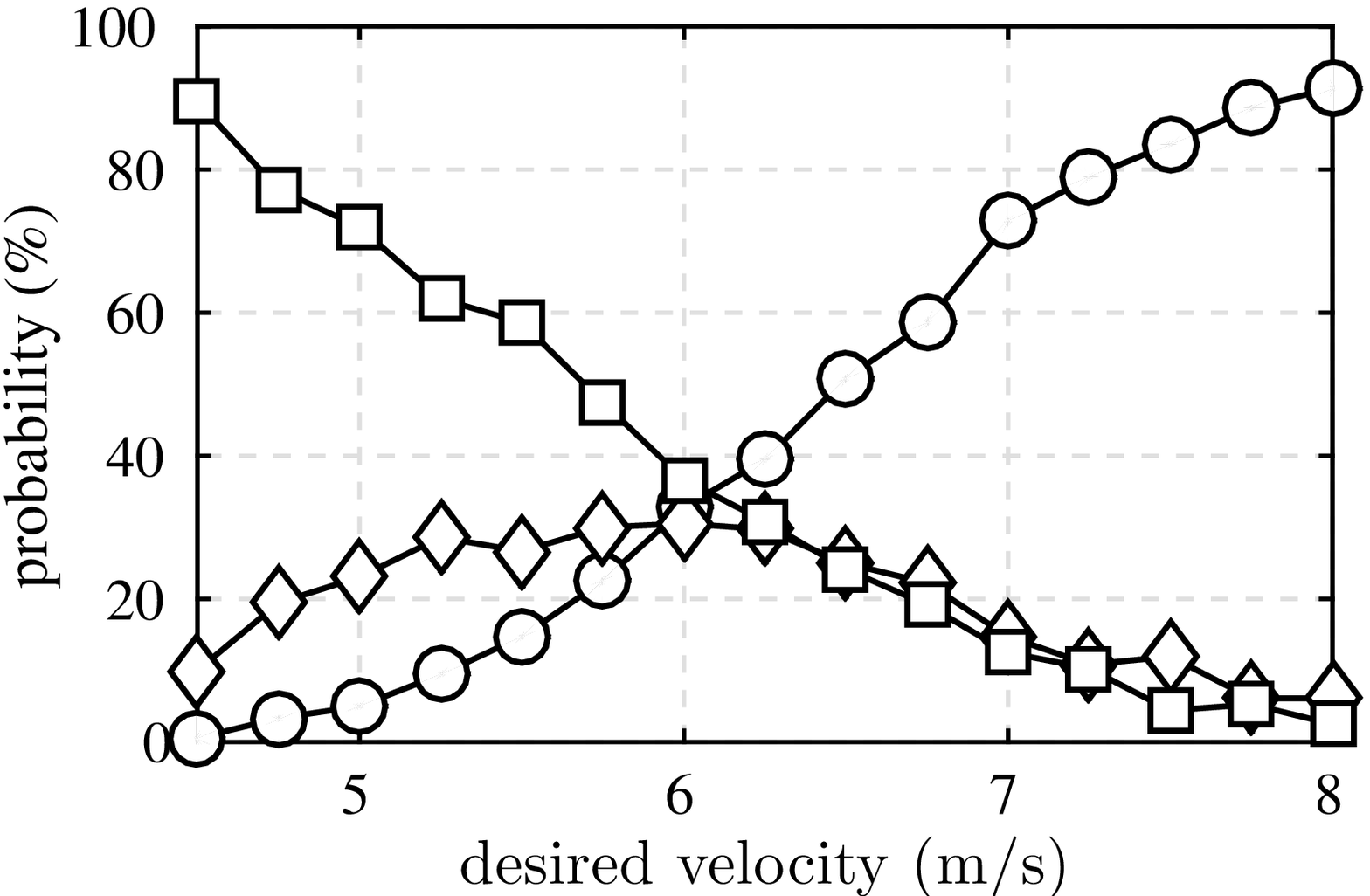}
}\hfill
\subfloat[ \label{fig:15}]{
\includegraphics[scale=0.4]{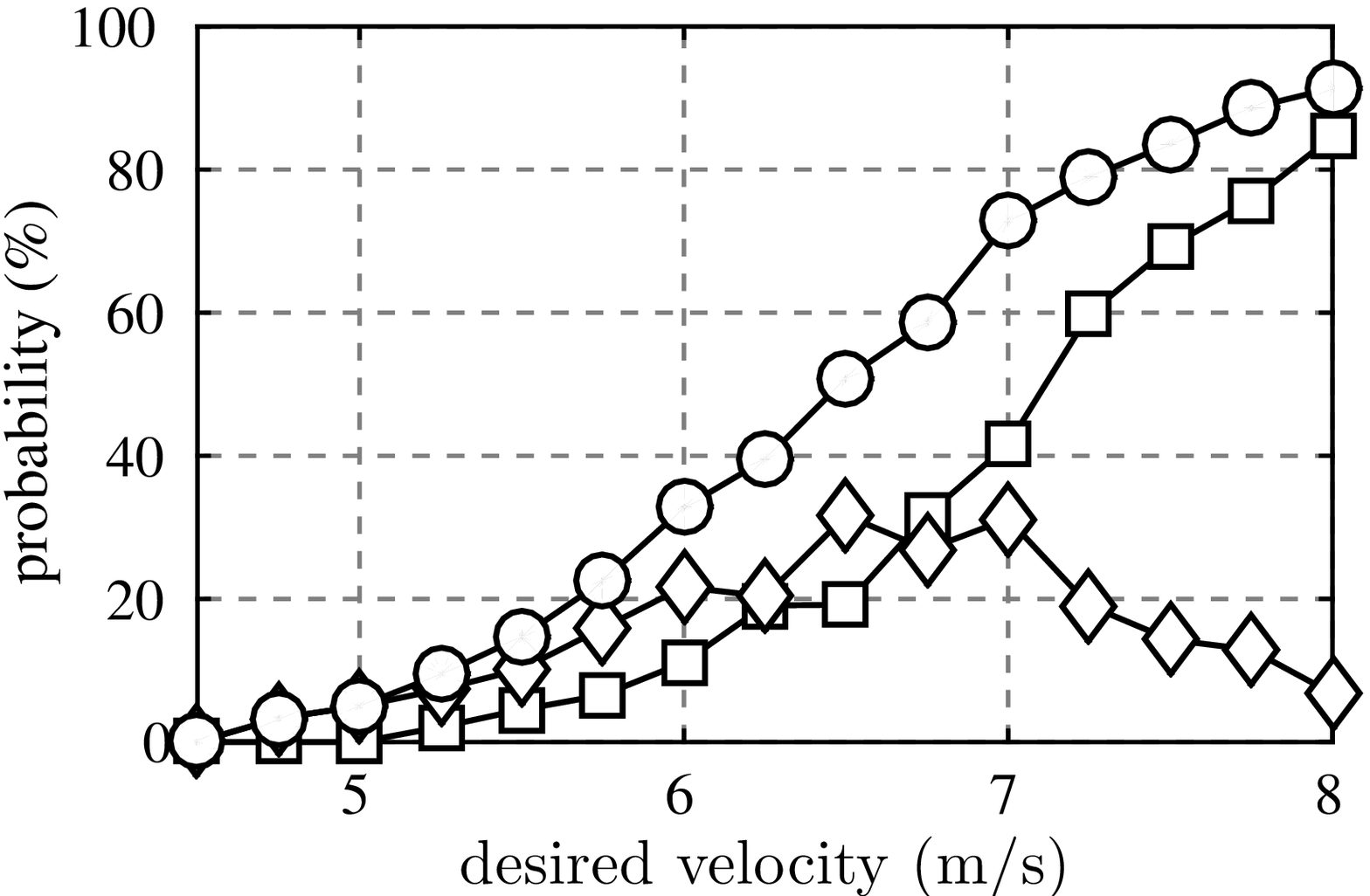}
}

\caption{\label{fig:types_probability}(a) Probability for each type of 
evacuation process vs. desired velocity (anxiety level). \mbox{$\bigcirc$ 
represent} the ``region 1'' processes (blocking). \mbox{$\diamondsuit$ 
represent} the ``region 2'' processes (narrow pathway). \mbox{$\Box$ represent} 
the ``region 3'' processes (wide pathway). (b) Contribution of each type of 
blocking to the total blocking probability of the exit vs. desired velocity 
(anxiety). \mbox{$\bigcirc$ represent} the blocking probability of the exit. 
$\diamondsuit$ represent the partial blocking probability of the exit. $\Box$ 
represent the total blocking probability of the exit. All the probabilities 
were computed over 360 realizations (both plots). The simulated time period 
was 300~s.} 
\end{figure*}

Fig.~\ref{fig:17} shows the probability of attaining any 
of the three evacuation situations (\textit{i.e.} regions 1, 2 or 3). We 
observe that the probability for the region 1 raises as the desired velocity 
increases, that is, as the individuals become more and more anxious. The region 
3 processes decrease along the same interval of $v_d$. But, the region 2 
situation achieves a maximum at 6~m/s. Notice that the three situations are 
equally likely only at 6~m/s. This is a nice anxiety level for equally sampling 
all 
the possible situations. \\

It becomes clear from Fig.~\ref{fig:17} that the desired 
velocity is somehow a control parameter for attaining any of the three possible 
situations. As the desired velocity is increased, the evacuation processes loci 
shown in Fig.~\ref{fig:4} move from ``region 3'' to ``region 1'', resembling a 
counterclockwise movement. However, this does not mean that ``region 2'' is an 
intermediate step between ``region 3'' and ``region 1''. As mentioned in 
Section~\ref{flux}, the probability of attaining ``region 2'' processes 
decreases as the simulation time increases. \\

The blocking configuration of the ``region 1'' situation also changes as the 
desired velocity increases. Fig.~\ref{fig:15} shows the \textit{total} and 
\textit{partial} blocking probability  as a function of the desired velocity. 
The \textit{total} blocking probability becomes relevant beyond $v_d=7\,$m/s 
with respect to the \textit{partial} blocking probability. This means that 
all the moving pedestrians will be locked due to a barrier surrounding the 
exit (for our simulation conditions).  \\

We conclude that the number of leaving pedestrians will depend on the desired 
velocity (\textit{i.e.} anxiety level) of the individuals, according to  our 
simulations (and for the current initial conditions). The desired velocity 
controls the probability of attaining any of the three possible situations, 
that is, the situations labeled as region 1, 2, or 3 (see Fig.~\ref{fig:4}). 
\\

\subsection{\label{pass_through} The passing-through scenario}

We now assume a different behavioural pattern for the moving pedestrians: 
they are able to pass-through unconscious (fallen) individuals in order to get 
out of the room. They no longer dodge the fallen pedestrians, since we assume 
that the passing-through is always possible, regardless of the additional 
difficulty that implies this new dynamic (see Section \ref{passoverforce} for 
details). \\

In \ref{corridor} we isolated a few pedestrians and studied the passing-through 
dynamics for the theoretical situations of a single and multiple lanes. We 
estimated the passing-through velocity $v_p$ and the relaxation time $\tau'$ 
following a specific criterion (see Section \ref{passoverforce} for details). 
Thus, \ref{corridor} provides the parameters used in this Section.   \\

\begin{figure*}
\subfloat[snapshot ($v_{p}$=0~m/s)\label{recinto_salto_vs_0_snapshot}]{
\vspace{-100mm} 
\includegraphics[scale=0.6]{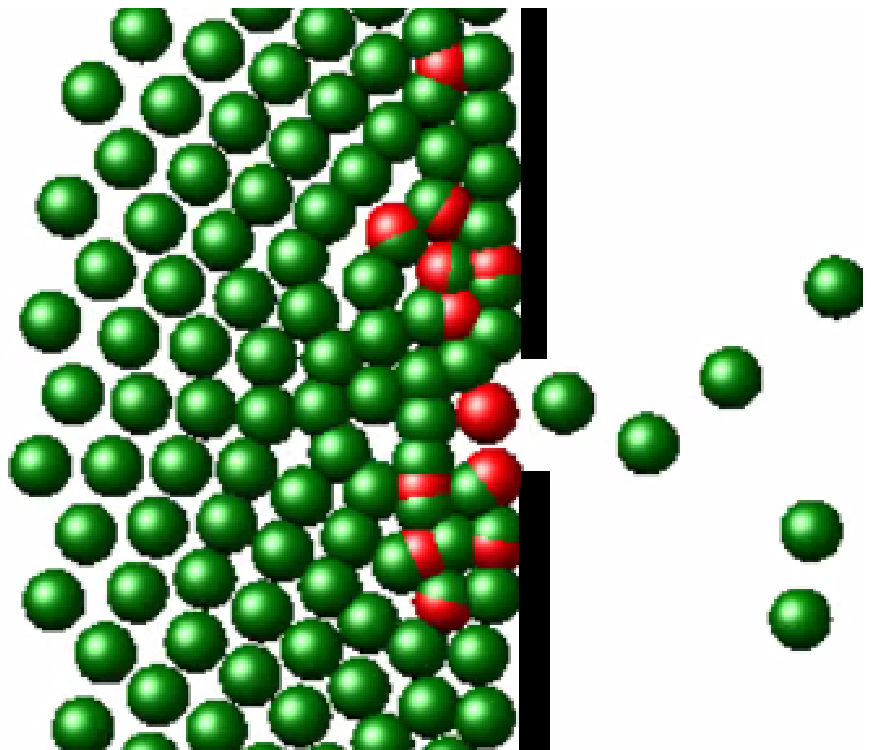}
}\hfill
\subfloat[passing-through ($v_{p}$=0~m/s)\label{recinto_salto_vs_0_delay}]{
\includegraphics[scale=0.4]{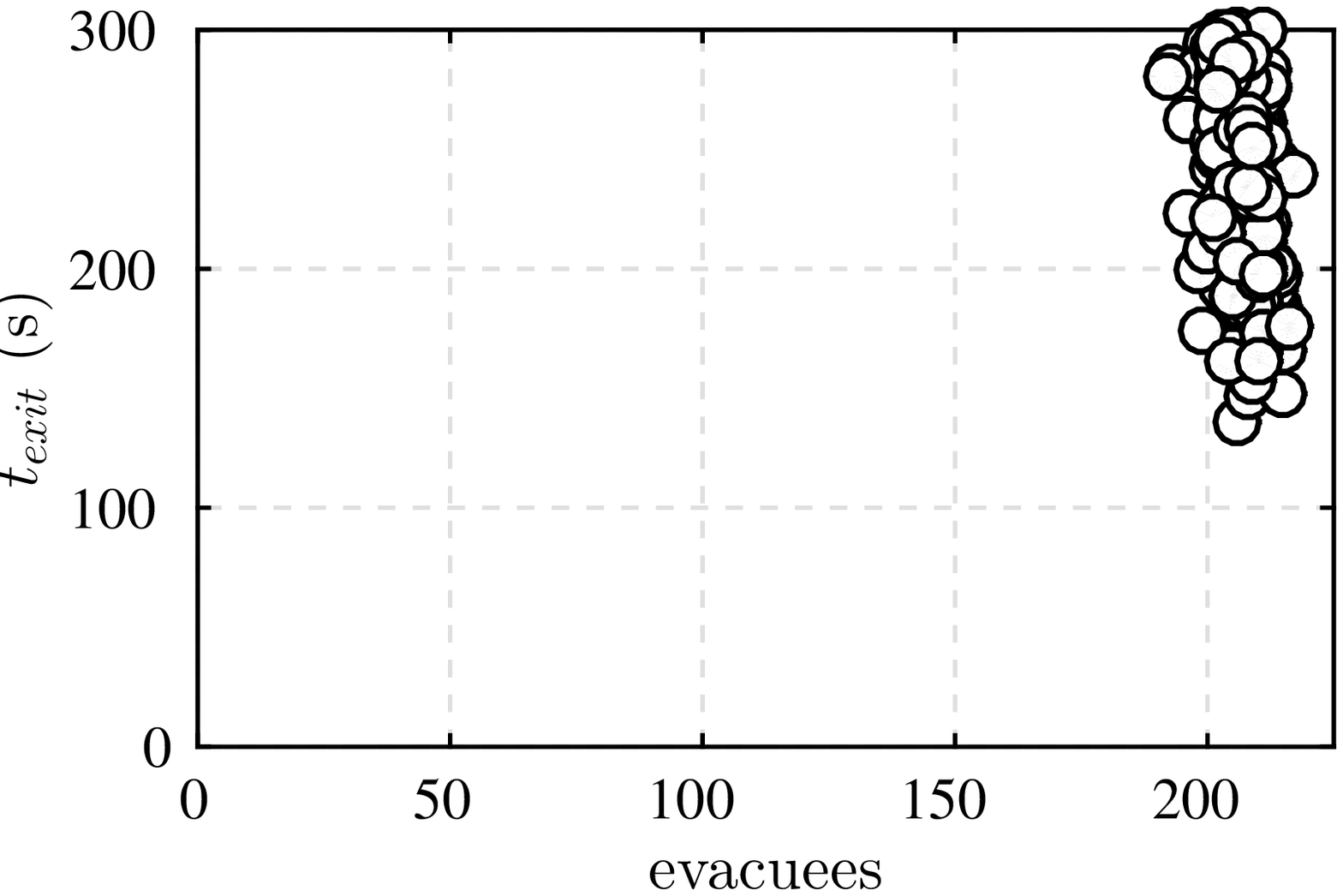}
}
\caption{(a) (color on-line only) Snapshot of an evacuation process at 100~s. 
Moving and fallen pedestrians are represented in green and red circles, 
respectively. The black lines represent the walls on the right of the room. The 
pedestrian's desired velocity was $v_d=6\,$m/s and the passing-through velocity 
was $v_p=0\,$m/s. (b) Leaving time $t_{exit}$ of the last evacuee as a function 
of the number of evacuees. Each circle represents an evacuation process (360 
processes are actually represented). All the evacuation processes were 
recorded along the first 300~s. The desired velocity was $v_d=6\,$m/s and the 
passing-through velocity was $v_p=0\,$m/s.\label{recinto_delays}}
\end{figure*}

Fig.~\ref{recinto_salto_vs_0_snapshot} shows a snapshot of an evacuation 
process in the passing-through scenario. The overlapping individuals are 
actually the ones passing through unconscious (fallen) pedestrians. Recall from 
Section~\ref{sec:Conditionfall} that no repulsive feelings due to fallen 
pedestrians actuate on the passing-through individuals. In this context, the 
moving pedestrians can manage to get out, and thus, the evacuation process can 
be fulfilled (except for the unconscious people). \\

The processes loci for the passing-through scenario is shown in 
Fig.~\ref{recinto_salto_vs_0_delay} for the desired velocity of $v_d=6\,$m/s 
and the passing-through velocity of $v_p=0\,$m/s. The null value of $v_p$ 
means that the passing-through individual experiences a moving difficulty 
such that his (her) willing vanishes. The passing-through pedestrian actually 
moves forward due to the individuals pushing from behind.\\

A quick comparison between Fig.~\ref{fig:4} 
and Fig.~\ref{recinto_salto_vs_0_delay} shows that switching from the dodging 
scenario to the passing-through one shifts the ``region 1'' and ``region 2'' 
loci to the ``region 3'' location. Therefore, we realize that the 
passing-through dynamic enhances the evacuation performance for those 
situations where dodging achieves a narrow pathway, or, some kind of blocking 
(\textit{i.e.} total or partial). \\

Notice that the ``region 1'' situation becomes relevant for 
$v_d>6\,$m/s, according to Fig.~\ref{fig:17}. This means that the evacuation 
enhancement will not be significant below this range. The same can be said 
about the ``region 2'' situation.\\ 

\begin{figure*}
\subfloat[ \label{presion_flujo_vd_6_vs_0}]{
\includegraphics[scale=0.4]{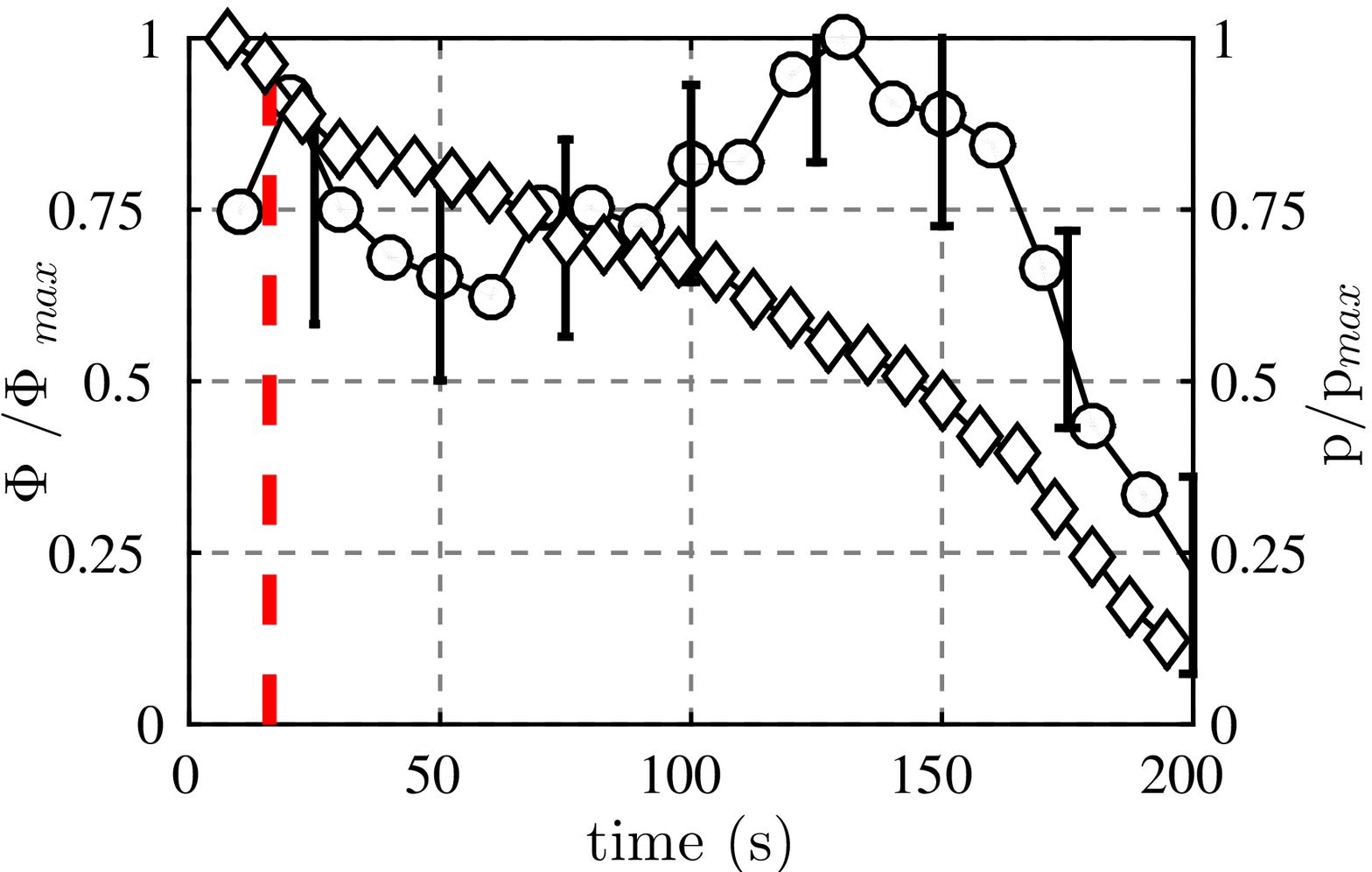}
}
\hfill
\subfloat[ \label{presion_evac_vd_6_vs_0}]{
\includegraphics[scale=0.4]{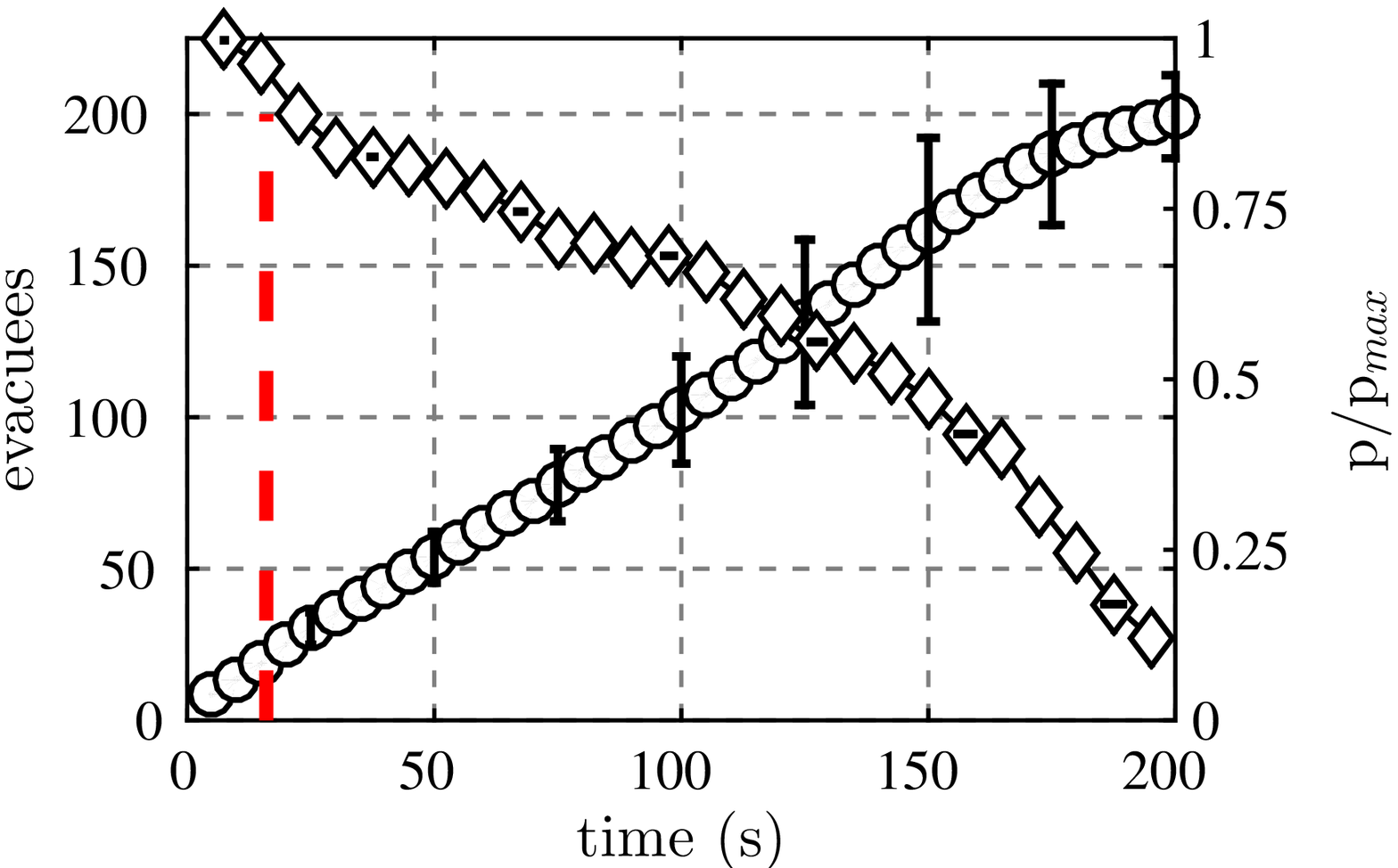}
}
\hfill
\subfloat[ \label{presion_flujo_vd_6_sin_caidos}]{
\includegraphics[scale=0.4]{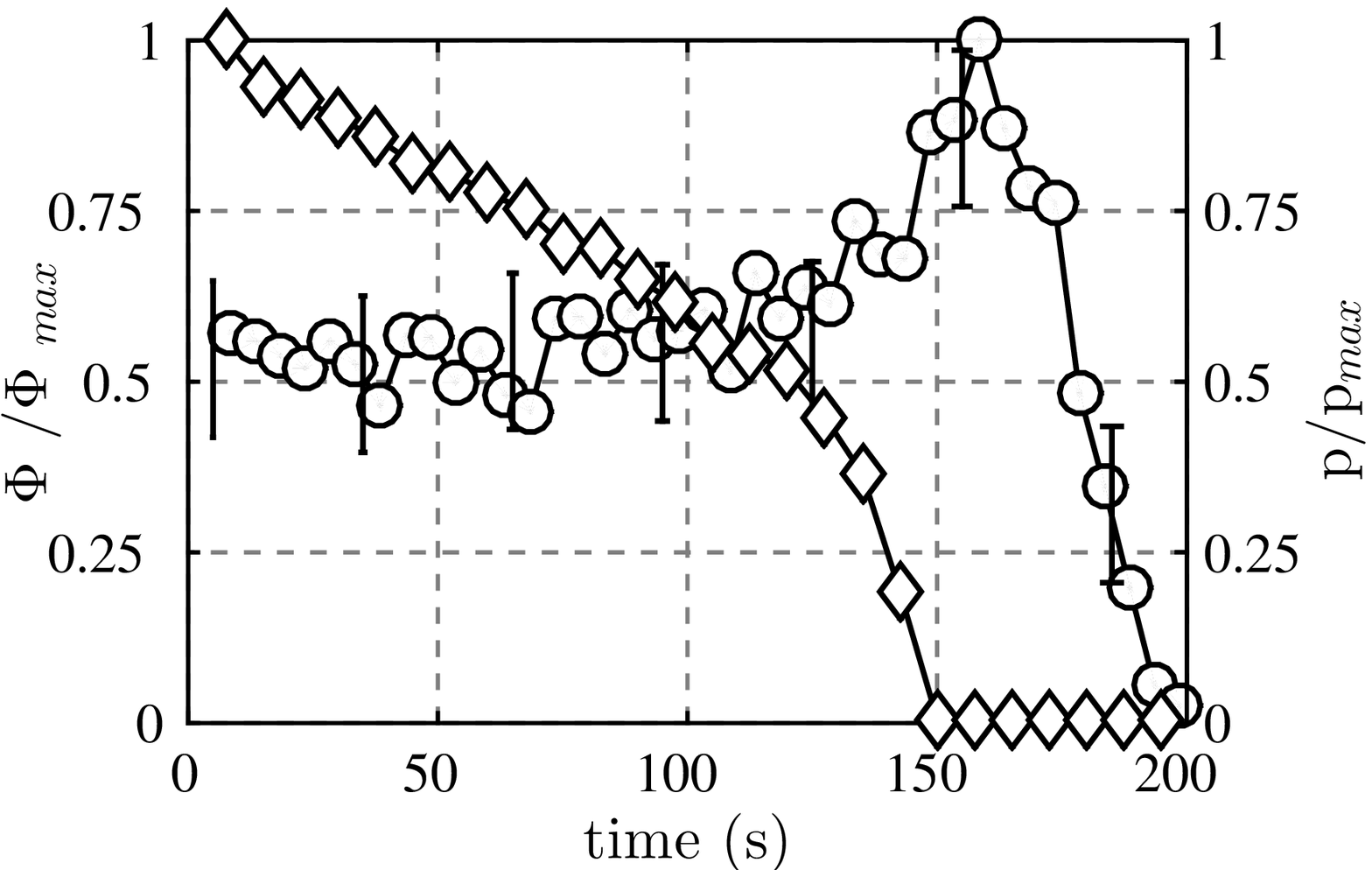}
}
\hfill
\subfloat[ \label{presion_evac_vd_6_sin_caidos}]{
\includegraphics[scale=0.4]{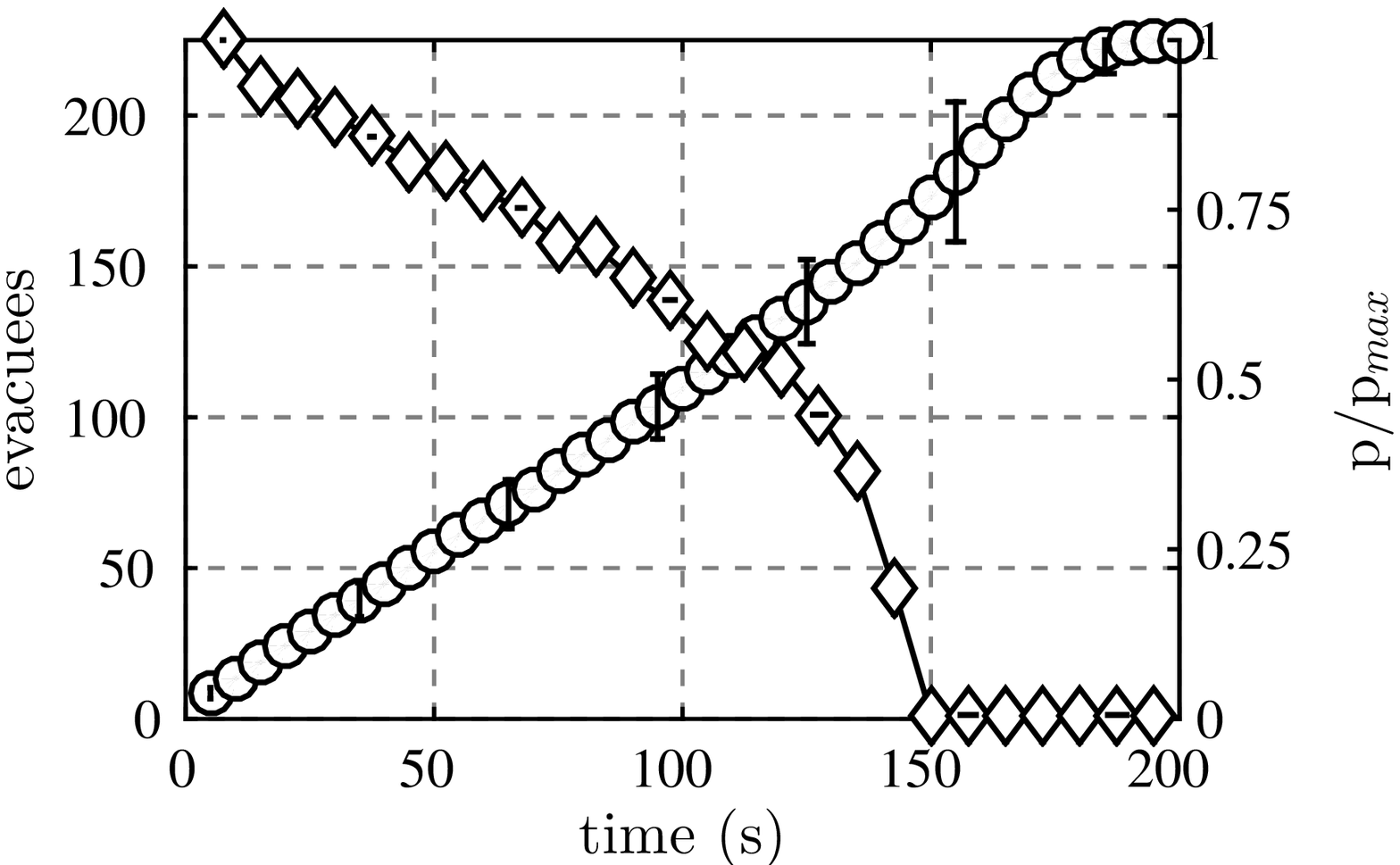}
}

\caption{\label{flujos_salto}(Color on-line only) The (a) and (b) plots 
correspond to the passing-through scenario with passing-through velocity  
$v_{p}$=0~m/s. The (c) and (d) plots correspond to the non-unconscious 
pedestrian scenario. The vertical red dashed line corresponds to 
the time stamp for the maximum number of fallen pedestrians per unit time (see 
Fig.~\ref{fallen_per_time}). For (a) and (c) plots: \mbox{$\bigcirc$ represent} 
the normalized evacuees flow rate ($\Phi /\Phi_{\textup{max}}$) and 
\mbox{$\diamondsuit$ represent} the normalized mean social pressure 
(p/p$_{\textup{max}}$) vs. time. For (b) and (d) plots: \mbox{$\Box$ represent} 
the number of evacuees and \mbox{$\diamondsuit$} represent the normalized mean 
social pressure (p/p$_{\textup{max}}$) vs. time.  Mean values were computed from 
50 realizations. Only the moving pedestrian contributed to the mean social 
pressure computation in (a) and (b) plots. The curves were normalized to have 
its maximum at unity. The error bars corresponds to $\pm\sigma$ (one standard 
deviation). The desired velocity was 
$v_d=6\,$m/s.} 
\end{figure*}

Since the passing-through dynamics improve the evacuation processes, we 
asked ourselves for the differences between this scenario and the one with 
non-unconscious pedestrians. That is, we investigated how similar  
could both scenarios be. Fig.~\ref{flujos_salto} exhibit the flow rates and 
mean pressures for both scenarios. \\

The pressure patterns in Fig.~\ref{presion_flujo_vd_6_vs_0} and 
Fig.~\ref{presion_flujo_vd_6_sin_caidos} correspond to the passing-through and 
non-unconscious scenarios, respectively.  Both patterns look very similar for 
the first 100~s (roughly, the first half of the process), but somehow 
differentiate beyond this interval. The  social pressure for the 
non-unconscious scenario decreases sharply until vanishing at 150~s. On the 
contrary,  the passing-through scenario does not vanish, but diminishes to a 
lower level. This level corresponds to the pressure on the pedestrians that are 
not able to get out, since their passing-through velocity is null ($v_p=0$).  \\

The flow rate patterns represented in Fig.~\ref{presion_flujo_vd_6_vs_0} and 
Fig.~\ref{presion_flujo_vd_6_sin_caidos} are also quite similar. The 
(normalized) flow rate for the passing-through scenario appears higher than the 
one for the non-unconscious scenario because the former does not present a 
sharp maximum close to 150~s as the latter. This is in agreement with the 
vanishing pressure shown in Fig.~\ref{presion_flujo_vd_6_sin_caidos}. That is, 
no individuals remain locked behind any unconscious pedestrian, and thus, 
pressure can be completely released.  \\

We conclude from Fig.~\ref{flujos_salto} that, although the passing-through and 
non-unconscious scenarios correspond to qualitatively different dynamics, the 
overall evacuation performance is quite similar for the null  
passing-through velocity ($v_p=0$). The only noticeable difference corresponds 
to those individuals that can not manage to leave the room because of
unconsciousness or null passing-through velocity $v_p=0$. \\

\begin{figure}
\begin{center}
 \includegraphics[scale=0.35]{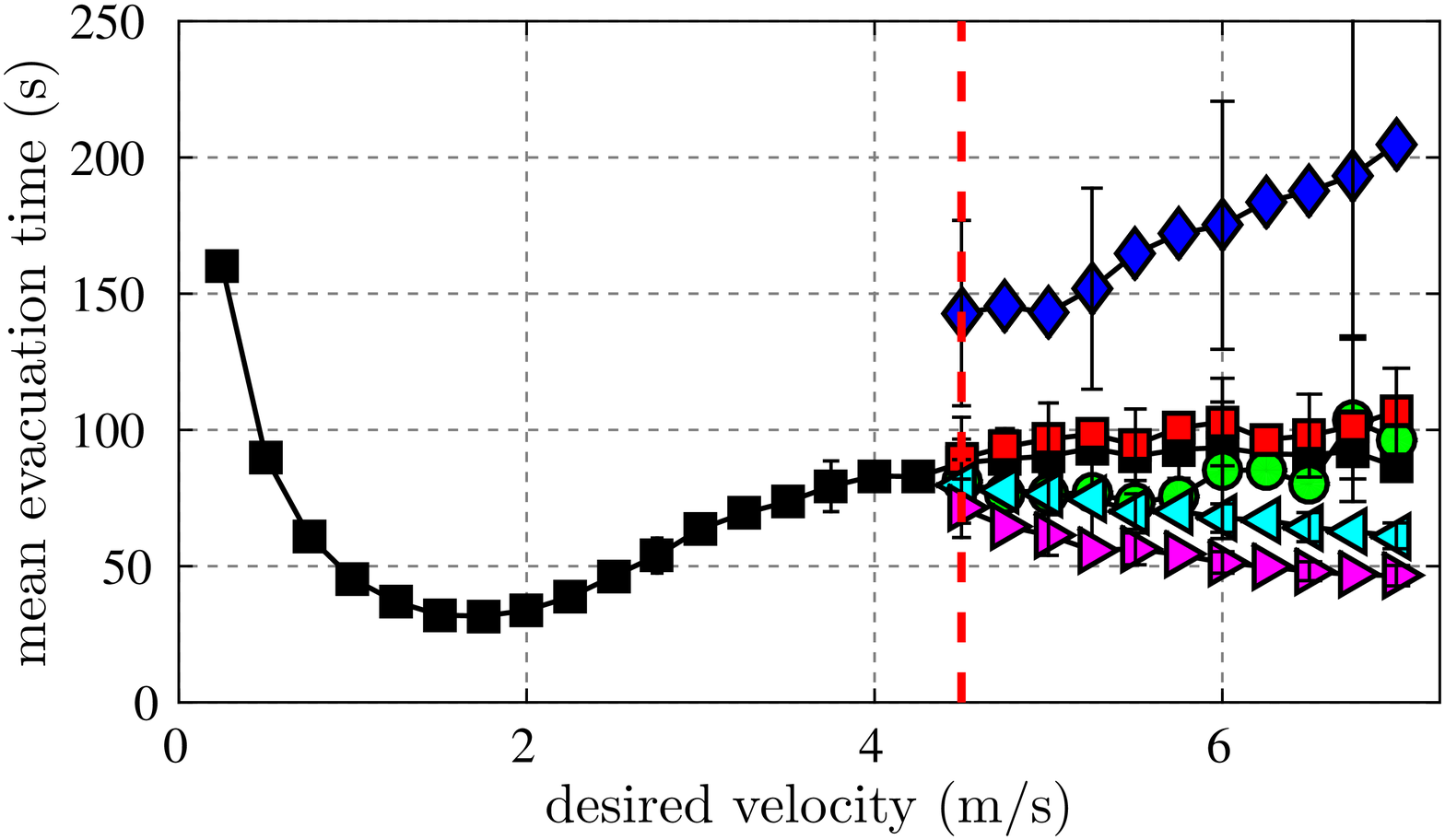}
   \end{center}
 \caption{(Color on-line only) Mean evacuation time for 100 individuals as a 
function of the desired velocity. 
  $\ \ $\protect\tikz\protect\draw[black,fill=black] (0,0) rectangle 
(0.3cm,0.3cm); 
 No fallen pedestrians (``faster is slower''); 
 \mydiamond{Blue} Narrow pathway; 
 \protect\tikz\protect\draw[black,fill=red] (0,0) rectangle (0.3cm,0.3cm); Wide 
pathway; 
 \protect\tikz\protect\draw[black,fill=green] (0,0) circle (.9ex);~v$_p$= 
1~m/s; 
 \mytriangleleft{cyan} v$_p$= 3~m/s; 
 \mytriangleright{magenta} v$_p$= 6~m/s; 
 \protect\hwplotB limit between those situations with no fallen 
(unconscious) individuals and thew ones with fallen pedestrians; mean values 
were computed from 30 realizations; the error bars corresponds to $\pm\sigma$ 
(one standard deviation).}
 \label{faster_is_slower_all} 
\end{figure}

We finally compared the overall performance for the different scenarios, as 
shown in Fig.~\ref{faster_is_slower_all}. The evacuation time in 
Fig.~\ref{faster_is_slower_all} corresponds to the time interval until 100 
individuals leave the room, in order to include the slow processes from the 
``region 2'' situation. Notice that the unconscious (fallen) scenario data 
points lie beyond $v_d=4.5\,$m/s since there are no fallen individuals for 
lower desired velocities. \\

According to Fig.~\ref{faster_is_slower_all}, the ``region 3'' situation does 
not actually improve the evacuation with respect to the non-unconscious scenario 
for the first 100 pedestrians leaving the room. But, the passing-through 
dynamics improves the evacuation performance for increasing values of $v_p$. 
Recall that $v_p$ regulates the difficulty experienced by the pedestrians 
passing through unconscious individuals. As $v_p$ increases, his (her) willing 
gets stronger because the degree of difficulty is supposed to diminish.  \\  

A brief summary for the evacuation performance in the two investigated contexts 
can be expressed as follows: the dodging scenario may worsen the 
evacuation performance with respect to the non-unconscious scenario, but, the 
passing-through scenario may improve the evacuation performance with respect to 
the same non-unconscious scenario. Both cases, the worsening or the 
enhancement, occur under certain conditions only. We were able to identify the 
desired velocity $v_d$ and the passing-through willing $v_p$ as two relevant 
control parameters for achieving this changes (for the setting mentioned in 
Section~\ref{numerical_geometry}). The worsening becomes noticeable for desired 
velocity $v_d>6\,$m/s. Besides, the enhancement becomes noticeable for 
passing-through willings $v_p>3\,$m/s.   \\

\section{\label{conclusions}Conclusions}

Our research focused on the high pressure scenarios during an emergency 
evacuation. Pressure is responsible for asphyxia and unconsciousness during 
the evacuation. People may fall, while others will further manage to escape. 
Two opposed scenarios are likely to happen: the moving pedestrians dodge the 
unconscious individuals, or, they manage to pass through them. In order to 
face these scenarios in the context of the ``social force model'', we 
\textit{hypothesized} that a ``passing-through'' force may be present or not, 
in order to attain either one scenario or the other. We stress that this 
is a \textit{first approach} to the aforementioned scenarios.\\

According to our simulations, unconsciousness is more likely to occur on the 
sides of the exit. For this reason, the unconscious (fallen) individuals not 
always block the exit completely. We arrived to the unexpected conclusion that 
neither the number of unconscious (fallen) pedestrians nor the falling rate 
of these individuals are relevant for the probability of blocking the exit. 
This conclusion holds for a fixed desired velocity $v_d=6\,$m/s. \\

We first focused on the dodging scenario. This scenario \textit{assumes} that 
moving pedestrians \textit{always} dodge the unconscious (fallen) 
individuals. As opposed to the ``passing-through'' scenario, the evacuation 
performance strongly depends on how the unconscious (fallen) individuals group 
into clusters.  Our research was able to distinguish between those situations 
that block the exit, and the ones where a free pathway remains open. The 
pathway width was found to be relevant for the evacuation performance. 
Therefore, three situations were well established: the blocking (totally or 
partially) situation, the narrow pathway  situation (roughly, one 
individual's diameter) and the wide pathway situation. The overall 
performance of the dodging scenario depended on the probability of attaining 
any of these three possible situations.  \\

We acknowledged that the evacuation process becomes interrupted after a short 
time period for the blocking situation. This is the worst situation, since 
many pedestrians get locked in the room because to the blocking clusters. On 
the contrary, if the grouping of unconscious (fallen) pedestrians allows a wide 
pathway to remain open, the dodging scenario does not show a significant 
worsening with respect to the lack of unconscious pedestrians. \\

The most interesting effect was captured for the narrow pathway situation. The 
moving pedestrians were only able to leave the room one after the other, in a 
stop-and-go process. This is a novel result and explains the significant 
slowing down that occurs for some processes in the dodging scenario.  \\

Our investigation on the dodging scenario explored a wide range of desired 
velocities, that is, we varied the anxiety level of the pedestrians. We 
specifically examined the range $4\,\mathrm{m/s}<v_d<8\,\mathrm{m/s}$. We 
concluded that the probability for the wide pathway situation was only relevant 
along the lower half of this range. Instead, the blocking situation became 
relevant for the upper half. The narrow path situation was relevant only 
around $v_d=6\,$m/s. All these conclusions showed that the desired velocity 
(or anxiety level of the pedestrians) is a control parameter for attaining any 
of the three possible situations. This is valid for the fixed initial 
conditions detailed in Section~\ref{numerical_geometry}.  \\  

We secondly focused on the ``passing-through'' scenario. Recall that we 
postulated the existence of a ``passing-through'' force in order to achieve a 
\textit{first approach} to this scenario. In this context, the pedestrian that 
passes through a fallen individual overcomes any blocking, although the 
difficulties, since the other pedestrians pushing from behind makes him move 
forward. Therefore, the overall evacuation performance improves with respect 
to the dodging scenario. Our investigation shows, however, that the 
passing-through willings need to surpass certain threshold 
(say, $v_p>3\,$m/s) for the improvement to become noticeable.\\

\section*{Acknowledgments}
C.O.~Dorso is a main researcher of the National Scientific and Technical 
Research Council (spanish: Consejo Nacional de Investigaciones Cient\'\i ficas y 
T\'ecnicas - CONICET), Argentina. G.A.~Frank is an assistant researcher of the 
CONICET, Argentina. F.E.~Cornes has degree in Physics. \\

\section*{References}

\bibliography{paper}

\appendix

\section{\label{corridor} Lanes of unconscious pedestrian}

This appendix examines the behaviour of one or more lanes of pedestrians 
willing to pass through unconscious (fallen) individuals. The 
``passing-through'' pedestrians move from left to right. The unconscious 
pedestrians are grouped in a compact cluster, located in the way of the 
``passing-through'' pedestrians. Two situations follow: the single lane 
situation and the multiple lane situation.  \\

\subsection{The single lane situation}

\begin{figure*}
\subfloat[Single lane \label{narrow corridor}]{
\vspace{-100mm} 
\includegraphics[scale=0.3]{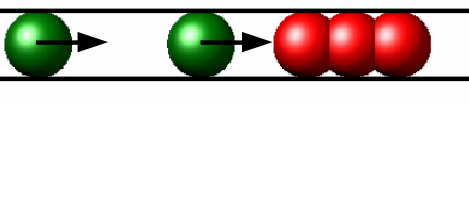}
}\hfill
\subfloat[Multiple lanes \label{wide corridor}]{
\includegraphics[scale=0.5]{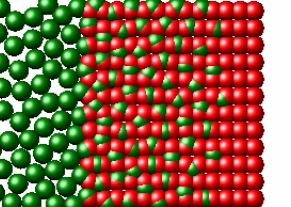}
}
\caption{(Color on-line only) Snapshots of different evacuation processes used 
to study the effects over the pedestrians that pass through the fallen 
individuals. Moving and fallen pedestrians are represent in green and red 
circles, respectively. The pedestrian's desired velocity was $v_d=6\,$m/s and 
the pass through velocity was $v_p=1\,$m/s. (a) Narrow corridor with two moving 
pedestrians and three fallen individuals. The arrows represent the direction of 
the desired velocity. The black lines represent the walls of the corridor. (b) 
Wide corridor (5.5~m of width) with 135 moving pedestrians and 155 fallen 
individuals. The moving direction is from left to right.\label{fig:corridor}} 
\end{figure*}

The most simple process that we can imagine corresponds to a single pedestrian 
passing through a small group of unconscious (fallen) individuals, as shown in 
Fig.~\ref{narrow corridor}. \\

\begin{figure}
\begin{center}
\includegraphics[scale=0.3]{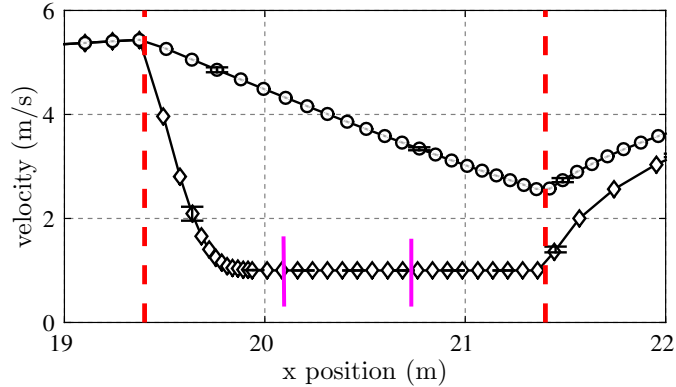}
\caption{Velocity of a pedestrian through a lane of three fallen individuals. 
\mbox{$\bigcirc$ represent} a reaction time of ${\tau}'$=0.5~s (corresponds to 
the basic social force model). \mbox{$\diamondsuit$ represent} a reaction time 
of ${\tau}'$=0.05~s. Mean values were computed from 30 realizations. The error 
bars corresponds to $\pm\sigma$ (one standard deviation). The vertical red 
dashed lines represent the initial and the ending positions of the fallen 
pedestrian's lane. The vertical magenta solid line represents the initial and 
the ending position of each fallen pedestrian. The individual moves free until 
the lane of fallen pedestrian. The pedestrian's desired velocity was set to 
$v_d=6\,$m/s, while the passing-through velocity was set to $v_p=1\,$m/s. 
\label{fig:pasillo_angosto_tau} }
\end{center}
\end{figure}

The ``passing-through'' pedestrian does not experience a social repulsive force 
due to the unconscious individuals, but the willing of passing through them 
$v_p$. Thus, solving Eq.~(\ref{eq_mov}) for this process arrives to 
the expressions

\begin{equation}
 \left\{\begin{array}{rcl}
         v & = & v_p+(v_d-v_p)\,e^{-(t-t_0)/\tau'} \\
           &   & \\
         x & = & x_0
-\tau'\,\bigg[v-v_d+v_p\,\ln\bigg(\displaystyle\frac{v_p-v}{v_p-v_d}\bigg)\bigg]
 \\
        \end{array}\right.\label{eqn:mov_appendix}
\end{equation}

\noindent for the initial conditions $x(t_0)=x_0$ and $v(t_0)=v_d$. The latter 
expresses that the pedestrian is moving freely before reaching the first 
unconscious (fallen) individual at position $x_0$ and time $t_0$. 
Fig.~\ref{fig:pasillo_angosto_tau} is a simulation of 
Eq.~\ref{eqn:mov_appendix} for two different values of $\tau'$. \\

Notice from Fig.~\ref{fig:pasillo_angosto_tau} that a passing-through 
relaxation time of $\tau'=0.5\,$s is too long for the pedestrian to reach the 
passing-through velocity $v_p$ within the distance of the first fallen 
individual (that is, $0.6\,$m). But, shrinking the relaxation time value to 
roughly 10\% accelerates the process, in order to reach $v_p$ within the 
expected distance. Thus, the value $\tau'=0.05\,$s is now meaningful, according 
to the definition given in Section~\ref{passoverforce}. Recall that this is a 
\textit{first approach} for the passing-through dynamics.\\  

The willing for passing through unconscious (fallen) pedestrians is regulated 
by the passing-through velocity $v_p$. There is currently no experimental value 
for $v_p$ in the literature to our knowledge. But, $v_p$ represents a 
slowing down in $v_d$ due to the difficulties of the ``passing-through'' 
context (see Section~\ref{passoverforce}). We fixed $v_p=1\,$m/s  in 
Fig.~\ref{fig:pasillo_angosto_tau} as a first example. Other possible values 
can 
be found in Section~\ref{pass_through}.  \\

\begin{figure}
\begin{center}
\includegraphics[scale=0.3]{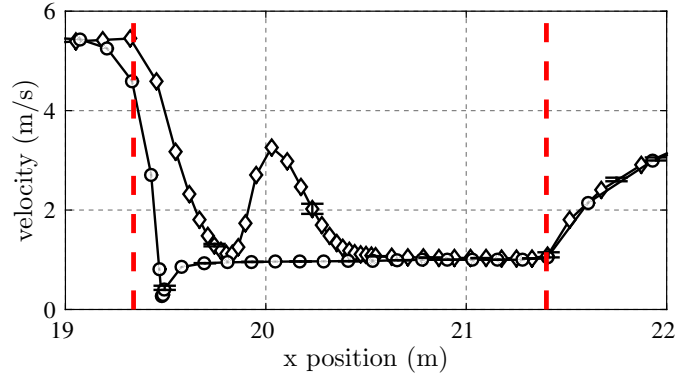}
\caption{Velocity of two pedestrians through the lane of three fallen 
individuals. The value of ${\tau}'$ was 0.05~s. $\diamondsuit$ represent the 
first individual to reach the lane of fallen pedestrians. $\bigcirc$ represent 
the second individual that reaches lane of fallen pedestrians (the left most 
moving pedestrian). Mean values were computed from 30 realizations. The error 
bars correspond to $\pm\sigma$ (one standard deviation). The vertical red dashed 
lines represent the initial and the ending positions of the lane of fallen 
pedestrians. The individuals moves freely until they reach the lane of fallen 
pedestrians. The pedestrian's desired velocity was $v_d=6\,$m/s and the 
passing-through velocity was $v_p=1\,$m/s. 
\label{fig:pasillo_angosto_2_vivos} }
\end{center}
\end{figure}

We further included a second pedestrian passing through the lane, as shown in   
 Fig.~\ref{narrow corridor}. Both moving pedestrians pass through the 
unconscious individuals, one after the other. 
Fig.~\ref{fig:pasillo_angosto_2_vivos} exhibits their velocity profiles as a 
function of the position.  \\

The velocity profile in Fig.~\ref{fig:pasillo_angosto_2_vivos} for the 
first pedestrian passing through the unconscious (fallen) individuals differs 
from the profile in Fig.~\ref{fig:pasillo_angosto_tau}. There is a maximum 
velocity immediately after the initial position of the lane (red line in  
Fig.~\ref{fig:pasillo_angosto_2_vivos}). This maximum corresponds to the 
pushing effect of the pedestrian behind him (her). Thus, our model for 
passing through unconscious individuals succeeds in capturing the effect of 
``pushing from behind''. Furthermore, this pushing effect allows the moving 
pedestrians to pass through the unconscious individuals even though 
$v_p=0\,$m/s. That is, if the passing-through pedestrian experiences a moving 
difficulty such that his (her) willing vanishes.  \\

Notice in Fig.~\ref{fig:pasillo_angosto_2_vivos} that the second pedestrians 
slows down immediately after entering the unconscious zone. This is the 
counterpart effect of ``pushing from behind''. \\

\subsection{The multiple lane situation}

\begin{figure*}
\subfloat[$v_{p}$=0~m/s\label{pasillo_ancho_v_media_vs_0}]{
\vspace{-100mm} 
\includegraphics[scale=0.38]{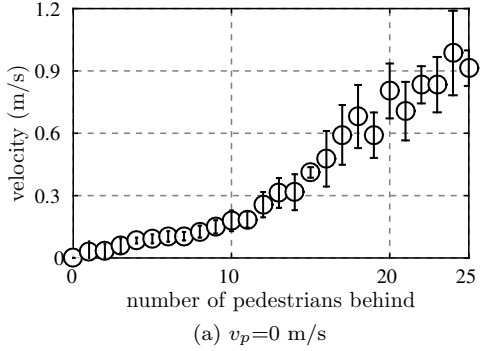}
}\hfill
\subfloat[$v_{p}$=1~m/s\label{pasillo_ancho_v_media_vs_1}]{
\includegraphics[scale=0.38]{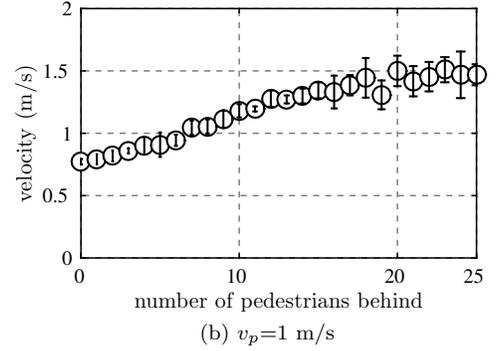}
}
\caption{Mean velocity per individual as a function of the number of 
pedestrians behind him (her). Mean values were computed from 30 realizations. 
The error bars corresponds to $\pm\sigma$ (one standard deviation). The 
pedestrian's desired velocity was 
$v_d=6\,$m/s.\label{fig:pasillo_ancho_v_media}} 
\end{figure*}

We introduced a multiple lane situation in order to deep into the ``pushing 
from behind'' and the ``slowing down'' effects. As shown in Fig.~\ref{wide 
corridor}, an arrangement of $12\times13$ unconscious (fallen) individuals was
placed at the right of 135 moving pedestrians. The moving pedestrians had the 
desire to go to the right. We simulated two situations: the pedestrian's 
passing-through velocity was null ($v_p=0\,$m/s), or, the passing-through 
velocity was $v_p=1\,$m/s. The former corresponds to pedestrians experiencing 
greater difficulties to surpass the unconscious individuals than the latter. 
Fig.~\ref{fig:pasillo_ancho_v_media} shows the mean velocity of the moving 
pedestrians as a function of the number of pushing people from behind (see 
caption for details).\\

Fig.~\ref{pasillo_ancho_v_media_vs_0} exhibits a null velocity if there are no 
other pushing pedestrians behind him (her). This is right, since the 
surpassing difficulties resembles a vanishing willing ($v_p=0\,$m/s). But, as 
more people push from behind, his (mean) velocity increases. For 25 pushing 
pedestrians, the mean velocity is close to 1~m/s. \\

Fig.~\ref{pasillo_ancho_v_media_vs_1} exhibits a mean velocity below 1~m/s if 
there are no pedestrians pushing from behind. This is less than $v_p$ and 
corresponds to the ``slowing down'' due to the pedestrians in front of him (that 
is, at the right of his current position). Notice that the ``slowing down'' 
diminishes as more people push from behind. At some point, both effects (the 
``slowing down'' and the ``pushing'') balance and the mean velocity becomes 
similar to $v_p=1\,$m/s. For 25 pushing pedestrians, the mean velocity of the 
passing-through individuals converges to 1.5~m/s. This value is in agreement 
with the measured velocity of the single individual in 
Fig.~\ref{fig:pasillo_angosto_2_vivos}. That is, from  
Fig.~\ref{fig:pasillo_angosto_2_vivos} we can expect a mean value between 1~m/s 
and 2~m/s. It also confirms that the ``slowing down'' is no longer relevant when 
25 individuals push from behind.   \\










\end{document}